\documentclass[12pt]{article}
\usepackage{amsmath}
\usepackage{amsfonts}
\usepackage{graphicx}
\usepackage[dvips]{epsfig}
\psfigdriver{dvips}
\usepackage{mathrsfs}

\begin{document}

%
\def\papertitlepage{\baselineskip 3.5ex \thispagestyle{empty}}
\def\preprinumber#1#2{\hfill \begin{minipage}{4.2cm}  #1
                 \par\noindent #2 \end{minipage}}
\renewcommand{\thefootnote}{\fnsymbol{footnote}}
\newcommand{\beq}{\begin{equation}}
\newcommand{\eeq}{\end{equation}}
\newcommand{\beqa}{\begin{eqnarray}}
\newcommand{\eeqa}{\end{eqnarray}}
\catcode`\@=11
\@addtoreset{equation}{section}
\def\theequation{\thesection.\arabic{equation}} 
\catcode`@=12
\relax
\newcommand{\Det}{\operatorname{Det}}
\newcommand{\Str}{\operatorname{Str}}
\newcommand{\tr}{\operatorname{tr}}

\newcommand{\ld}{\lambda}
\newcommand{\ks}{k\hskip-1.2ex /}
%
%
%
\papertitlepage
\setcounter{page}{0}
\preprinumber{KEK-TH-1121}{hep-th/0612080}
\baselineskip 0.8cm
\vspace*{2.0cm}
\begin{center}
{\large\bf Progress toward the Determination of \\
Complete Vertex Operators for 
The I$\hspace{-.1em}$IB Matrix Model}
\end{center}
\vskip 4ex
\baselineskip 1.0cm
\begin{center}
           {Yoshihisa Kitazawa${}^{1,2}$, Shun'ya Mizoguchi${}^{1,2}$ 
and Osamu Saito${}^{1,3}$} 
\\
\vskip 1em
       ${}^1${\it High Energy Accelerator Research Organization (KEK)} \\
       \vskip -2ex {\it Tsukuba, Ibaraki 305-0801, Japan} \\
\vskip 1em
       ${}^2${\it Department of Particle and Nuclear Physics} \\
       \vskip -2ex {\it The Graduate University for Advanced Studies} \\
       \vskip -2ex {\it Tsukuba, Ibaraki 305-0801, Japan}\\
\vskip 1em
      ${}^3$ {\it Institute for Cosmic Ray Research, University of Tokyo} \\
        \vskip -2ex {\it Kashiwa 277-8582, Japan}
\end{center}
\vskip 5ex
%
\baselineskip=3.5ex
\begin{center} {\bf Abstract} \end{center}
We report on progress in determining the complete form of vertex operators 
for the I$\hspace{-.1em}$IB matrix model. The exact expressions are obtained for those 
emitting massless I$\hspace{-.1em}$IB supergravity fields up to sixth order in the 
light-cone superfield, in which the conjugate gravitino and conjugate 
two-form vertex operators are newly determined.
We also provide a consistency check by computing the kinematical factor 
of a four-point graviton amplitude in a D-instanton background.
We conjecture that the low-energy effective action of the I$\hspace{-.1em}$IB matrix model 
at large $N$ is given by tree-level supergravity coupled to the vertex operators.

\vspace*{\fill}
\noindent
December 2006
\newpage
\renewcommand{\thefootnote}{\arabic{footnote}}
\setcounter{footnote}{0}
\setcounter{section}{0}
\baselineskip = 0.6cm
\pagestyle{plain}

\section{Introduction}
Realizing string theory as a matrix model is an efficient way for investigating 
its nature beyond perturbation. Among others, the I$\hspace{-.1em}$IB matrix model is
in a sense the simplest proposal, and there are several good reasons to believe that it is 
a natural framework for type I$\hspace{-.1em}$IB string theory. The first direct link between them is that 
the former may be regarded as a matrix regularization of the latter \cite{IKKT}. 
Another suggestive fact is that, defined as a zero-dimensional reduced model
of the maximal supersymmetric Yang-Mills theory, the I$\hspace{-.1em}$IB matrix model can be thought 
of as an effective theory of D-instantons.  Since its has twice as many supercharges as  
the original Yang-Mills, it must also contain a graviton. Indeed the analysis based on the 
one-loop effective action has already revealed an interaction due to a graviton 
exchange \cite{IKKT}. Furthermore, the continuum limit of the loop equation was shown to 
reproduce the type I$\hspace{-.1em}$IB light-cone superstring field theory \cite{FKKT}. 

One of the issues of the I$\hspace{-.1em}$IB matrix model is how it describes the coupling to the 
supergravity multiplet. The first study of this question was undertaken by one of the 
present authors \cite{Kitazawa}, who constructed a set of vertex operators 
for the massless supergravity multiplet to leading order in momentum $k$ 
by repeatedly applying the ${\cal N}=2$ SUSY transformations to the straight 
Wilson line operator. There was also found a set of  ``wave functions"  which are 
dual to the vertex operators and linearly transform under the ${\cal N}=2$ SUSY 
transformations.

To fully determine the complete form of them is a hard task.
In \cite{ITU} it was noticed that the supersymmetric Wilson loop operators 
were useful tools in determining the complete form of vertex operators.
In general, Wilson loops are basic gauge invariant objects in any gauge theory.
In particular, in the I$\hspace{-.1em}$IB matrix model,
they describe creations and annihilations of 
fundamental strings \cite{IKKT,FKKT}, of which 
the lowest excitations are the massless fields of I$\hspace{-.1em}$IB supergravity 
multiplet. Therefore it is natural that 
Wilson loop operators  come into play in the construction of 
vertex operators in the I$\hspace{-.1em}$IB matrix model. 
Expanding a supersymmetric Wilson loop operator as a polynomial of the ``mean field"
fermion variable $\ld$, which may be identified as the light-cone superfield variable, 
exact  vertex operators have been systematically derived \cite{ITU}  up to fourth 
order in $\ld$.

If one goes beyond that order, one is faced with the enormous task of algebras 
and Fierz arrangements.
In this paper, to simplify the computations, we develop some alternative ways of 
deriving them, as we will explain shortly. Combining information obtained through 
these as well as other available means, we determine the complete form of 
vertex operators up to {\em sixth} order in $\ld$. 
The expressions for  vertex operators emitting a conjugate gravitino ($\Psi_\mu^c$) and 
a conjugate two-form field ($B^c_{\mu\nu}$) are the new results.
To compute the remaining two most complicated ones of ${\cal O}(\ld^7)$ and ${\cal O}(\ld^8)$ 
(corresponding to the conjugate dilatino and the conjugate dilaton) still require another 
enormous amount of labor, and they are not determined so far. The work is in progress. 

The contents of this paper are as follows. In section 2, we briefly review the general concept of 
vertex operators in the I$\hspace{-.1em}$IB matrix model and the idea of how to compute them. 
In section 3, we describe the method adopted in \cite{ITU} as well as our improved 
ones which simplifies the calculation. In section 4,  we present 
the new expressions for the conjugate gravitino and 
conjugate two-form vertex operators. We also display  other vertex operators 
known so far for completeness. In section 5 a consistency check is given by computing 
the kinematical factor of the four-point graviton amplitude. Section 6 is devoted to 
the conclusions. In two Appendices we collect useful formulas and  summarize  our results.

\section{Vertex operators for the I$\hspace{-.1em}$IB matrix model}
The action of the I$\hspace{-.1em}$IB matrix model is \cite{IKKT}
\beqa
S&=&-\frac14 tr{[} A_\mu,A_\nu{]}^2 -\frac12 tr\left(
\bar\psi \Gamma^\mu{[} A_\mu, \psi{]}
\right),
\eeqa
where $A_{\mu}$ and $\psi$ are $N \times N$ Hermitian matrices 
which transform as a $D=10$ 
Lorentz vector and a Majorana-Weyl spinor. 
Vertex operators in the I$\hspace{-.1em}$IB matrix model are given by functions of them, 
and are characterized by the properties:
(i) They describe linear couplings to the background fields.
(ii) They are related with each other by the supersymmetry transformations.
%
 Let $V_{i}(A,\psi)$ be the vertex operator associated with the background field $f_i$ 
which is any of the members of type I$\hspace{-.1em}$IB supergravity multiplet. 
Then the property 
(i) implies that the interaction terms are given by
\begin{equation}
       S_{int}=\sum_{i} V_{i}(A,\psi)f_i,
\end{equation}
and (ii) asserts that the following equation holds \cite{DNP}:
   \begin{equation}
       \sum_iV_{i}(\delta A,\delta \psi)f_i = \sum_i V_{i}(A,\psi)\delta f_i,
    \label{variation}
    \end{equation}
where $\delta$ is a supersymmetry variation. 
The relation (\ref{variation}) ensures the super invariance of correlation functions 
\begin{equation}
 W(f_i)\equiv \langle \sum_i V_{i}(A,\psi)f_i \rangle,
\end{equation}
and, in principle, determines the form of vertex operators completely \cite{Kitazawa}.
%
%
%
Indeed, 
the vertex operators for all members of the I$\hspace{-.1em}$IB supergravity multiplet 
were 
determined in this way to leading order in momentum $k$ \cite{Kitazawa}.
The derivation of the complete form   
becomes, however, more complicated as the vertex operator comes to include more fermions.




%
    
The authours of \cite{ITU} have developed, by utilizing a supersymmetric Wilson loop 
operator,  a more systematic way of determining the exact form of vertex operators 
in the I$\hspace{-.1em}$IB matrix model. 
%
%
 %
%
%
%
Let us focus on the operator 
\begin{equation}
    w(A,\psi;\lambda)=e^{\bar{\lambda}Q_1}tr e^{ik\cdot A}e^{-\bar{\lambda}Q_1},
\end{equation}
where $\lambda$ is a Majorana-Weyl spinor and $Q_1$ is one of the ${\cal N}=2$ 
SUSY generators of the I$\hspace{-.1em}$IB matrix model
\begin{eqnarray}
    \bar{\epsilon_1}Q_1 
    &=&i(\bar{\epsilon_1}\Gamma_{\mu}\psi)\frac{\delta}{\delta A_{\mu}}
    -\frac{i}{2}F_{\mu\nu}\Gamma^{\mu\nu}
     \epsilon_1\frac{\delta}{\delta \psi},
     \label{homogeneous}\\
\bar{\epsilon}_2 Q_2 &=&\epsilon_2 \frac{\delta}{\delta \psi},
\label{Q2}
\end{eqnarray}
where $F_{\mu\nu}=[A_{\mu},A_{\nu}]$.
$w(A,\psi;\lambda)$ is the simplest supersymmetric Wilson loop operator first introduced in \cite{Hamada}.
One of the nice properties of $w(A,\psi;\lambda)$ is that it transforms under the 
${\cal N}=2$ SUSY as \cite{ITU}
\begin{eqnarray}
     [\bar{\epsilon}Q_1,w(A,\psi;\lambda)]&=&\epsilon\frac{\delta}{\delta \lambda}w(A,\psi;\lambda),
\label{supersym1}\\
     { [}\bar{\epsilon}Q_2,w(A,\psi;\lambda){]}&=&(\bar{\epsilon}\ks \lambda)w(A,\psi;\lambda).
\label{supersym2}
\end{eqnarray}
In other words, the operations (\ref{homogeneous})(\ref{Q2}) acting on the space of functions of matrices 
$A_\mu$ and $\psi$ amounts to the operations 
\begin{eqnarray}
       \delta^{(1)}&=&\epsilon \frac{\delta}{\delta \lambda},
        \label{SUSY1}\\
       \delta^{(2)}&=&\bar\epsilon\ks \lambda
         \label{SUSY2}
\end{eqnarray}
on the space of polynomials of $\ld$. 
Since (\ref{SUSY1})(\ref{SUSY2}) realize the ${\cal N}=2$ SUSY algebra, 
one may construct a representation in this space. 
In the following we take $k^2=0$, then the irreducible subspace is spanned by 
monomials of $\ld$ of at most degree eight. Such a polynomial can be regarded
as a light-cone superfield for the massless typeI$\hspace{-.1em}$IB supergravity multiplet \cite{GSB}.

The basic strategy of deriving vertex operators adopted by \cite{ITU} is as follows :
We first find a set of homogeneous polynomial of $\ld$ such that each of them corresponds 
to some supergravity field and transforms as a linearized SUSY multiplet by the operations 
(\ref{SUSY1})(\ref{SUSY2}). Such a polynomial is called a {\it wave function}.
We expand the supersymmetric Wilson loop $w(A,\psi;\ld)$ as a polynomial of $\lambda$
in terms of wave functions
\begin{equation}
    w(A,\psi;\lambda)=\sum_{i} V_{i}(A,\psi)f_i(\lambda). 
\label{wexpansion}
\end{equation}
One immediately finds
   \begin{equation}
        \sum_{i}[\bar{\epsilon}Q_j,V_i(A,\psi)]f_i(\lambda) = \sum_i V_i(A,\psi)\delta^{(j)}f_i(\lambda)
     \end{equation}
  for $j=1,2$.
Thus the supersymmetric Wilson loop realizes the relation
(\ref{variation}), and
%
the coefficient of each wave function can be identified as the corresponding vertex 
operator. 
The fermionic variable $\ld$ may be regarded as an isolated eigenvalue of the matrix $\psi$ 
representing the effect of the background as a ``mean field" \cite{ITU,ISTU}. Indeed the SUSY 
transformations for such a single eigenvalue are generated by (\ref{SUSY1})(\ref{SUSY2}) 
if the off-diagonal interactions are neglected. Thus we see that the wave functions for the 
external fields are realized as condensations of particular spinor eigenvalues of the matrix 
model.

\section{The methods}
In this section we will show how we determine the form of vertex operators in some 
detail.
    We begin by rewriting the supersymmetric Wilson loop as
\begin{equation}
  w(A,\psi;\lambda)=tr e^{ik\cdot A +\Sigma_{n=1}^{8}G_n},
\label{supersymmetric_Wilson_loop_G}
\end{equation}
where $G_{n}$ is defined by
\begin{eqnarray}
   G_n =\frac{1}{n!}[\bar{\lambda}Q_1,\cdots[\bar{\lambda}Q_1,ik\cdot A]\cdots]  \ \ (n \ {\rm commutators}),
\end{eqnarray}
and hence contains $n\ \lambda$'s. 
The sum  in the exponent of (\ref{supersymmetric_Wilson_loop_G})
terminates at $G_8$
because $\ld$ has  only eight independent components.
Each term can be evaluated as 
\begin{eqnarray}
   G_0&=& ik\cdot A ,\\
   G_1&=& \bar{\psi}\ks \lambda, \\
   G_2&=&\frac{i}{4}b^{\mu\nu}[A_{\mu},A_{\nu}], \\
   G_3&=& -\frac{1}{3!}b^{\mu\nu}[\bar{\lambda}\Gamma_{\mu},A_{\nu}],\\
   G_4&=&\frac{1}{4!}\left(\frac{i}{2}b^{\mu\nu}(\bar{\lambda}\Gamma_{\mu\rho\sigma}\lambda)[[A^{\rho},A^{\sigma}],A_{\nu}]
           -ib^{\mu\nu}[\bar{\lambda}\Gamma_{\mu}\psi,\bar{\lambda}\Gamma_{\nu}\psi]\right),\\
    G_5&=&-\left(\frac{1}{5!}{b^{\mu\nu}(\bar{\lambda}\Gamma_{\mu\rho\sigma}\lambda)
            [[\bar{\lambda}\Gamma^{\rho}\psi,A^{\sigma}],A_{\nu}]+\frac{3}{2}b^{\mu\nu}(\bar{\lambda}\Gamma_{\mu\rho\sigma}
    \lambda) [[A^{\rho},A^{\sigma}],\bar{\lambda}\Gamma_{\nu}\psi]}\right), \ \ \  \ \  \\
\vdots\nonumber 
\end{eqnarray}
Expanding (\ref{supersymmetric_Wilson_loop_G}) and collecting terms with 
the same powers of $\lambda$, we get
\begin{eqnarray}
     w(A,\psi;\lambda)
                      &=& Stre^{ik\cdot A}
      \bigg[ 1+G_1 +\left(\frac{1}{2!}(G_1)^2 +G_2\right) +\left( \frac{(G_1)^3}{3!}+G_1\cdot G_2 +G_3\right)
    \nonumber \\
   &&\hskip -2em+ \left( \frac{(G_1)^4}{4!}+\frac{1}{2}(G_1)^2\cdot G_2 +\frac{1}{2}(G_2)^2+G_1 \cdot G_3 +G_4 \right)
      \nonumber \\
    &&\hskip -2em+  \left( \frac{(G_1)^5}{5!}+\frac{1}{3!}(G_1)^3\cdot G_2 +\frac{1}{2}G_1\cdot (G_2)^2+\frac{1}{2}(G_1)^2\cdot G_3
       +G_2\cdot G_3 +G_1\cdot G_4 +G_5 \right)\nonumber \\
    &&\hskip -2em+\cdots  \bigg],
\end{eqnarray}
where ``$Str$" means the symmetrized trace
whose definitions and some properties are 
given in the appendix. 
If we let $i$ in  (\ref{wexpansion}) denote the power of $\ld$ (but see below for $i=4$), 
we obtain
\begin{eqnarray}
    V_0(A,\psi)f_0(\lambda) &=& Str e^{ik\cdot A},\label{v_0}\\
    V_1(A,\psi)f_1(\lambda) &=& Str G_1 e^{ik\cdot A},\label{v_1}\\
     V_2(A,\psi)f_2(\lambda) &=& Str \left( \frac{1}{2!}(G_1)^2 +G_2\right)e^{ik\cdot A},\\
    V_3(A,\psi)f_3(\lambda) &=& Str \left( \frac{(G_1)^3}{3!}+G_1\cdot G_2 +G_3\right)e^{ik\cdot A},\\
&&\vdots \nonumber \\
     V_n(A,\psi)f_n(\lambda) &=& Str\left( \frac{(G_1)^n}{n!}+\cdots +G_n\right)e^{ik\cdot A},\label{method1} \\
    &&  \vdots \nonumber 
\end{eqnarray}
If we rearrange the right hand side in terms of wave functions,  
we obtain corresponding vertex operators \cite{ITU}. 
Note that there are two independent wave functions (graviton and self-dual four-form field) 
at $i=4$. Therefore the terms of ${\cal O}(\ld^4)$ split into a linear combination of them; each coefficient is a 
vertex operator in this case. 

Although this method offers a systematic derivation, it  requires 
too many calculations for orders higher than four. 
In order to reduce the amount of labor in the derivation, we have developed the following two 
improved methods. The first one uses a different expansion of the supersymmetric Wilson loop as
%
\begin{eqnarray}
    w(A,\psi;\lambda)&=&e^{\bar{\lambda}Q_1}tr e^{ik\cdot A} e^{-\bar{\lambda}Q_1} \nonumber \\
                 &=& Str \bigg( e^{ik\cdot A}+[\bar{\lambda}Q_1,e^{ik\cdot A}]\nonumber \\
             &&+\frac{1}{2!}[\bar{\lambda}Q_1,[\bar{\lambda}Q_1,e^{ik\cdot A}]]
                 +\cdots +\frac{1}{8!}[\bar{\lambda}Q_1,\cdots [\bar{\lambda}Q_1,e^{ik\cdot A}]\cdots ]\bigg).\ \ \ \  
\label{new_expansion}
\end{eqnarray}
Again, the
sum terminates at ${\cal O}(\lambda^8)$. 
From ({\ref{new_expansion}})
we obtain
\begin{eqnarray}
    V_n(A,\psi)f_n(\lambda)&=& Str\frac{1}{n!}[\bar{\lambda}Q_1,\cdots [\bar{\lambda}Q_1,e^{ik\cdot A}]\cdots]\nonumber \\
        &=& Str \frac{1}{n!}\left( i(\bar{\lambda}\Gamma_{\mu}\psi)\frac{\delta}{\delta A_{\mu}}
                       -\frac{i}{2}F_{\mu\nu}\Gamma^{\mu\nu}\lambda\frac{\delta}{\delta\psi}\right)^n e^{ik\cdot A}.
\end{eqnarray}
%
This equation relates the ${\cal O}(\lambda^{n})$ term with the ${\cal O}(\lambda^{n-1})$ term as
\begin{equation}
      V_n(A,\psi)f_n(\lambda)=\frac{1}{n}\left( i(\bar{\lambda}\Gamma_{\mu}\psi)\frac{\delta}{\delta A_{\mu}}
                       -\frac{i}{2}F_{\mu\nu}\Gamma^{\mu\nu}\lambda\frac{\delta}{\delta\psi}\right)V_{n-1}(A,\psi)f_{n-1}(\lambda).
\label{method2}
\end{equation}
Therefore, once we know the exact expression for $V_{n-1}(A,\psi)$,  we can determine 
$V_n(A,\psi)$ by using this relation. Note that the operator in the right hand side 
is the homogeneous supersymmetry transformation (\ref{homogeneous}) with parameter $\lambda$. Thus, $V_n(A,\psi)$ is determined by supersymmetry. This is close to the original method 
adopted in \cite{Kitazawa}, but what is new here is that we can now use the precise 
expressions of wave functions obtained in \cite{ITU} as we present in the next subsection.

The second alternative method uses another recursion relation \cite{LittleIIB} 
\begin{equation}
   \bar{\epsilon}\frac{\delta}{\delta \psi}V_n(A,\psi)f_n(\lambda)
=(\bar{\epsilon}\ks \lambda)V_{n-1}(A,\psi)f_{n-1}(\lambda),
\label{method3}
\end{equation}
which can easily be shown by
comparing the ${\cal O}(\lambda^n)$ terms on the left and right hand sides of  
(\ref{supersym2}).
Using this formula, we can also determine the form of $V_{n}(A,\psi)$ if 
$V_{n-1}(A,\psi)$ is known. 
In this case $V_{n}(A,\psi)$ is transformed
by inhomogeneous supersymmetry transformation(\ref{Q2}).

The advantage of the first method is that we do not need to calculate the polynomial 
of $G_i$'s, which becomes longer as the power of $\ld$ gets higher. 
The disadvantage is that it is not easy to reexpress the polynomials of $\ld$ in terms of 
 wave functions.  On the other hand, the advantage of the second method is that
the right hand side can readily be written in terms of the next wave function by using 
the SUSY transformation formula. 
The disadvantage is that it is not a trivial problem to integrate the left hand side by $\psi$.
Anyway, by combining information obtained in various ways above, we have successfully 
obtained new expressions for the vertex operators of ${\cal O}(\ld^5)$ and ${\cal O}(\ld^6)$, as 
we now show. 

%

\section{The explicit forms of vertex operators}
We will now display our results. The expressions for the conjugate gravitino and 
conjugate two-form vertex operators are new, while
other vertex operators known so far are also shown for completeness. 
\subsection{Wave functions}
The wave functions corresponding to the massless I$\hspace{-.1em}$IB  supergravity
multiplet have already obtained \cite{Kitazawa, ITU}\footnote{Signs are 
corrected in (\ref{wavefnPsimuc})$\sim$(\ref{wavefnPsic}).}:
%
%
%
\begin{eqnarray}
 \Phi(\lambda) &=&1 \\
  \tilde{\Psi}(\lambda) &=& \ks \lambda \\
  B_{\mu\nu}(\lambda)&=&-\frac{1}{2}b_{\mu\nu} \\
  \Psi_{\mu}(\lambda)&=&\frac{i}{24}\Gamma^{\alpha}\ks \lambda b_{\alpha\mu}\\
  h_{\mu\nu}(\lambda)&=&\frac{1}{96}{b_{\mu}}^{\rho}b_{\rho\nu} \\
  A_{\mu\nu\rho\sigma}(\lambda) &=& -\frac{i}{32 \cdot (4!)^2}b_{ [ \mu\nu}b_{\rho\sigma]} \nonumber \\
                                  &=& -\frac{i}{4 \cdot (4!)^2}(b_{\mu\nu}b_{\rho\sigma}+b_{\mu\rho}b_{\sigma\nu}
  +b_{\mu\sigma}b_{\nu\rho})\\
   \Psi^{c}_{\mu}(\lambda) &=& -\frac{i}{4\cdot 5!}\Gamma_{\rho}\ks\lambda b^{\rho\sigma}b_{\sigma\mu}
   \label{wavefnPsimuc}\\
   B^{c}_{\mu\nu}(\lambda) &=& \frac{1}{6!}b_{\mu\rho}b^{\rho\sigma}b_{\sigma\nu} \\
  \tilde{\Psi}^{c} (\lambda)&=& \frac{1}{8!}\Gamma^{\mu\nu}\ks \lambda b_{\mu\rho}b^{\rho\sigma}b_{\sigma\nu}\\
    \Psi^{c}(\lambda) &=& -\frac{1}{ 8 \cdot 8!}{b_{\mu}}^{\nu}{b_{\nu}}^{\lambda}{b_{\lambda}}^{\sigma}{b_{\sigma}}^{\mu}.
    \label{wavefnPsic}
\end{eqnarray}
One may check the following SUSY transformations
\begin{eqnarray}
        \delta\Phi &=&  \bar{\epsilon}_2\tilde{\Phi} \\
        \delta\tilde{\Phi} &=& \ks \epsilon_1 \Phi -\frac{i}{24}\Gamma^{\mu\nu\rho}\epsilon_2 H_{\mu\nu\rho} \\
        \delta B_{\mu\nu} &=& -\bar{\epsilon}_1\Gamma_{\mu\nu}\tilde{\Phi}
                +2i (\bar{\epsilon}_2\Gamma_{[\mu}\Psi_{\nu]}+k_{\mu}\Lambda_{\nu]})\\
         \delta \Psi_{\mu} &=& \frac{1}{24\cdot 4} [9\Gamma^{\nu\rho}\epsilon_1H_{\mu\nu\rho}
                              -\Gamma_{\mu\nu\rho\sigma}\epsilon_1H^{\nu\rho\sigma}]
                            +\frac{i}{2}\Gamma^{\nu\rho}k_{\rho}h_{\mu\nu}\epsilon_2 \nonumber \\
                           &&  +\frac{i}{4\cdot 5!}\Gamma^{\rho_1 \cdot \cdot \cdot \rho_5}\Gamma_{\mu}\epsilon_2F_{\rho_1\cdot \cdot \cdot \rho_5}
                             +k_{\mu}\xi \\
       \delta h_{\mu\nu} &=& -\frac{i}{2}\bar{\epsilon}_1\Gamma_{(\mu}\Psi_{\nu)}
                              +\frac{i}{2}\bar{\epsilon}_2\Gamma_{(\mu}\Psi^{c}_{\nu)} +k_{(\mu}\xi_{\nu)} \\
       \delta A_{\mu\nu\rho\sigma} &=& -\frac{1}{(4!)^2}\bar{\epsilon}_1\Gamma_{[\mu\nu\rho}\Psi_{\sigma]}
                             -\frac{1}{(4!)^2}\bar{\epsilon}_2\Gamma_{[\mu\nu\rho}\Psi_{\sigma]}+k_{[\mu}\xi_{\nu\rho\sigma]}\\
       \delta \Psi^c_{\mu} &=& -\frac{i}{2}\Gamma^{\nu\rho}k_{\rho}h_{\mu\nu}\epsilon_1 +\frac{i}{4\cdot 5!}\Gamma^{\rho_1\cdot \cdot \cdot \rho_5}
                                \Gamma_{\mu}\epsilon_1 F_{\rho_1\cdot \cdot \cdot \rho_5} \nonumber \\
                            && +\frac{1}{24\cdot 4} [9\Gamma^{\nu\rho}\epsilon_2H^c_{\mu\nu\rho}-{\Gamma_{\mu}}^{\nu\rho\sigma}\epsilon_2H^c_{\nu\rho\sigma}]
                                +k_{\mu}\xi^c \\
       \delta B^c_{\mu\nu} &=& 2i (\bar{\epsilon}_1\Gamma_{[\mu}\Psi^c_{\nu]}+k_{[\mu}\Lambda^c_{\nu]})-\bar{\epsilon}_2\Gamma_{\mu\nu}\tilde{\Phi}^c\\
       \delta \tilde{\Phi}^c &=& -\frac{i}{24}\Gamma^{\mu\nu\rho}\epsilon_1H^c_{\mu\nu\rho} +\ks \epsilon_2 \Phi^c \\
       \delta \Phi^c &=& \bar{\epsilon}_1\tilde{\Phi}^c,
\end{eqnarray}
where $\xi,\xi_{\mu},\xi_{\mu\nu\rho}$ and $\Lambda_{\mu}$ are gauge parameters.




\subsection{Vertex operator for the dilaton $\Phi$ \cite{Kitazawa,ITU}}
    The dilaton vertex operator $V^{\Phi}$ is simply given by 
    (\ref{v_0}) :
\begin{equation}
   V^{\Phi}(A,\psi)=Str e^{ik\cdot A},
\end{equation}
where $k^2=0$. Since there is no operator other than $e^{ik\cdot A}$ in the symmetrized trace, "Str" is equivalent to the ordinary trace. 
     
\subsection{Vertex operator for dilatino $\tilde{\Phi}$ \cite{Kitazawa,ITU}}
     The dilatino vertex operator $V^{\tilde{\Phi}}$ is read off from the term with a single $\lambda$ (\ref{v_1}) as 
\begin{equation}
      V^{\tilde{\Phi}}(A,\psi)\tilde{\Phi}(\lambda)= Str e^{ik\cdot A}G_1=Stre^{ik\cdot A}\bar{\psi}\ks \lambda
=Stre^{ik\cdot A}\bar{\psi}\tilde{\Phi}(\lambda).
\end{equation}
We obtain 
\begin{equation}
   V^{\tilde{\Phi}}(A,\psi)=Stre^{ik\cdot A}\bar{\psi}.
\end{equation}
The symmetrized trace is identical to ordinary trace.

\subsection{Vertex operator of antisymmetric tensor field $B_{\mu\nu}$ \cite{Kitazawa,ITU}}
 The vertex operator for the antisymmetric tensor $B_{\mu\nu}$ can be obtained from the terms 
 containing two $\lambda$'s in the expansions of the supersymmetric Wilson loop.
Using (\ref{method2}), we have
\begin{eqnarray}
   V^B_{\mu\nu}(A,\psi)B^{\mu\nu}(\lambda)&=& Str \frac{1}{2}\left( (\bar{\lambda}\Gamma_{\mu}\psi)\frac{\delta}{\delta A_{\mu}}
     -\frac{i}{2}F_{\mu\nu}\Gamma^{\mu\nu}\lambda \frac{\delta}{\delta \psi}\right)V^{\tilde{\Phi}}(A,\psi)\tilde{\Phi}(\lambda)
      \nonumber \\
           &=& Str \frac{1}{2}\left( (\bar{\lambda}\Gamma_{\mu}\psi)\frac{\delta}{\delta A_{\mu}}
     -\frac{i}{2}F_{\mu\nu}\Gamma^{\mu\nu}\lambda \frac{\delta}{\delta \psi}\right)
              (\bar{\psi}\ks \lambda)e^{ik\cdot A}\nonumber \\
        &=& Str \left( \frac{1}{2}(\bar{\psi}\ks \lambda)\cdot(\bar{\psi}\ks \lambda)+\frac{i}{4}F_{\mu\nu}
                                          b^{\mu\nu}\right)e^{ik\cdot A}\nonumber \\
        &=& Str e^{ik\cdot A}\left(\frac{1}{16}\bar{\psi}\cdot \ks \Gamma_{\mu\nu}\psi-\frac{i}{2}F_{\mu\nu}\right)B_{\mu\nu}(\lambda).
\end{eqnarray} 
Hence the vertex operator for the antisymmetric tensor field is given by \cite{Kitazawa,ITU}
\begin{equation}
     V^B_{\mu\nu}(A,\psi) =Str e^{ik\cdot A}\left(\frac{1}{16}\bar{\psi}\cdot \ks \Gamma_{\mu\nu}\psi-\frac{i}{2}F_{\mu\nu}\right).
\end{equation}

The vertex operator satisfies
\begin{equation}
    k^{\mu}V^{B}_{\mu\nu}(A,\psi)=0.
\end{equation}
This implies that the coupling of the vertex operator with the background field \\$V^{B}_{\mu\nu}(A,\psi)B^{\mu\nu}(\lambda)$ is 
gauge invariant.

\subsection{Vertex operator for gravitino $\Psi_{\mu}$\cite{ITU}}
   The third order terms give the gravitino $\Psi_{\mu}$ vertex operator 
   \begin{equation}
     V^{\Psi}_{\mu}(A,\psi) =Str e^{ik\cdot A}\left( -\frac{i}{12}(\bar{\psi}\cdot 
             \ks \Gamma_{\mu\nu}\psi)-2F_{\mu\nu}\right)\cdot \bar{\psi}\Gamma^{\nu} .
\label{v_gravitino}
\end{equation}
It can be shown that (\ref{v_gravitino}) satisfies (\ref{method3}):
\begin{equation}
     \epsilon \frac{\delta}{\delta \psi}V^{\Psi}_{\mu}(A,\psi) \Psi^{\mu}(\lambda)
            =V^{B}_{\mu\nu}(A,\psi)(\bar{\epsilon}\ks \lambda)B^{\mu\nu}(\lambda).
\label{method3_3}
\end{equation}
Indeed, the left hand side is evaluated as follows
\begin{eqnarray}
     \epsilon \frac{\delta}{\delta \psi}V^{\Psi}_{\mu}(A,\psi) \Psi^{\mu}(\lambda)&=&
          Stre^{ik\cdot A}\left( -\frac{i}{6}k^{\rho}(\bar{\epsilon}\Gamma_{\mu\nu\rho}\psi)\cdot\bar{\psi}\Gamma^{\nu}
          \right.\nonumber\\
          &&\left.~~~~~~~~~~~~
                            -\frac{i}{12}k^{\rho}(\bar{\psi}\cdot\Gamma_{\mu\nu\rho}\psi)\bar{\epsilon}\Gamma^{\nu}
                               -2F_{\mu\nu}\bar{\epsilon}\Gamma^{\nu}\right)\Psi^{\mu}(\lambda)\nonumber \\
     &=& Str e^{ik\cdot A}\left( -\frac{i}{4}(\bar{\psi}\cdot\ks \Gamma_{\mu\nu}\psi)\bar{\epsilon}\Gamma^{\nu}\Psi^{\mu}(\lambda)
          -2F_{\mu\nu}\bar{\epsilon}\Gamma^{\nu}\Psi^{\mu}(\lambda)\right),
\end{eqnarray}
whereas the right hand side becomes
\begin{eqnarray}
    V^{B}_{\mu\nu}(A,\psi)(\bar{\epsilon}\ks \lambda)B^{\mu\nu}(\lambda) &=&
         Str e^{ik\cdot A}\left(\frac{1}{16}\bar{\psi}\cdot\ks \Gamma_{\mu\nu}\psi-\frac{i}{2}F_{\mu\nu}\right)
              2i\bar{\epsilon}\Gamma^{[\mu}\Psi^{\nu]}(\lambda).
\end{eqnarray}
Thus we have established the equation (\ref{method3_3}).

This gravitino vertex operator is shown to satisfy
  \begin{equation}
    k^{\mu}V^{\Psi}_{\mu}(A,\psi)=0.
   \label{gauge_invariance_3}
  \end{equation}
The first term of $V^{\Psi}_{\mu}(A,\psi)$ trivially satisfies this relation, while the second term is calculated as 
\begin{eqnarray}
      k^{\mu}V^{\Psi}_{\mu}(A,\psi)&=&Str e^{ik\cdot A}2i[ik\cdot A,A_{\nu}]\bar{\psi}\Gamma^{\nu}\nonumber \\
                  &=& Str [e^{ik\cdot A},A_{\nu}]2i \bar{\psi}\Gamma^{\nu}\nonumber \\
                   &=& -Str e^{ik\cdot A}2i [\bar{\psi},A_{\nu}]\Gamma^{\nu}\nonumber \\
                    &=&0.
\end{eqnarray}
In the last line we have used the equation of motion for $\psi$:
\beqa
\Gamma^\mu{[}A_\mu, \psi {]}=0.
\label{equation_of_motion_for_psi}
\eeqa 


\subsection{Vertex operators for graviton $h_{\mu\nu}$ 
                                                    and fourth-rank antisymmetric tensor field $A_{\mu\nu\rho\sigma}$
                                                    \cite{ITU}}

   The next terms with four $\lambda$ 's give the vertex operators for the graviton $h_{\mu\nu}$ and 
the fourth-rank antisymmetric tensor $A_{\mu\nu\rho\sigma}$. In order to derive them, we use (\ref{method2}) with $n=4$:
\begin{eqnarray}
    V^h_{\mu\nu}(A,\psi)h^{\mu\nu}(\lambda)+V^A_{\mu\nu\rho\sigma}(A,\psi)A^{\mu\nu\rho\sigma}(\lambda) 
       \phantom{yyyyyyyyyyyyyyyyyyyyyy}\nonumber \\
           = \frac{1}{4}\left( i(\bar{\lambda}\Gamma_{\mu}\psi)\frac{\delta}{\delta A_{\mu}}
                       -\frac{i}{2}F_{\mu\nu}\Gamma^{\mu\nu}\lambda\frac{\delta}{\delta\psi}\right)
           V^{\Psi}_{\mu}(A,\psi)\Psi^{\mu}(\lambda).
\end{eqnarray}
In evaluating the right hand side we use many Fierz identities and properties of symmetrized trace.
Here we only write down the final results. The vertex operator for the graviton is given by
 \begin{eqnarray}  
V^{h}_{\mu\nu}&=&Str e^{ik\cdot A} \bigg(-\frac{1}{96}k^{\rho}k^{\sigma}\left(\bar{\psi}\cdot 
{\Gamma_{\mu\rho}}^{\beta}\psi\right)\cdot\left(   \bar{\psi}\cdot 
\Gamma_{\nu \sigma\beta} \psi \right) \nonumber \\
&&-\frac{i}{4}k^{\rho}\bar{\psi}\cdot \Gamma_{\rho\beta(\mu}\psi \cdot {F_{\nu )}}^{\beta}
+\frac{1}{2}\bar{\psi}\cdot \Gamma_{(\mu}[A_{\nu )}, \psi]+2{F_{\mu}}^{\rho}\cdot F_{\nu\rho} \bigg).
\end{eqnarray}
The vertex operator for the fourth-rank antisymmetric tensor field is given by
\begin{eqnarray}
V^{A}_{\mu\nu\rho\sigma}&=& Str e^{ik\cdot A} \bigg( \frac{i}{8\cdot 4!}k_{\alpha}k_{\gamma} 
(\bar{\psi}\cdot {\Gamma_{[\mu\nu}}^{\alpha}\psi)\cdot (\bar{\psi}\cdot {\Gamma_{\rho\sigma]}}^{\gamma}\psi)
 +\frac{i}{3}\bar{\psi} \cdot \Gamma_{[\nu\rho\sigma}[\psi,A_{\mu]}]\nonumber \\
&&+\frac{1}{4}F_{[\mu\nu}\cdot (\bar{\psi}\cdot {\Gamma_{\rho\sigma]}}^{\gamma}\psi)k_{\gamma}
-iF_{[\mu\nu}\cdot F_{\rho\sigma]}\bigg).
\label{vertex_operator_A}
\end{eqnarray}

  These vertex operators satisfy the conservation laws
\begin{eqnarray}
     k^{\mu}V^h_{\mu\nu}(A,\psi)&=&0 ,\\
     k^{\mu}V^A_{\mu\nu\rho\sigma}(A,\psi)&=& 0 .
\end{eqnarray}
The first term of the graviton vertex operator trivially satisfies this equation, while the second term is 
evaluated as 
\begin{eqnarray}
     Str e^{ik\cdot A}\bigg( -\frac{1}{4}k^{\rho}(\bar{\psi}\cdot \Gamma_{\rho\beta\nu}\psi)[ik\cdot A,A^{\beta}]\bigg)
   &=& Str e^{ik\cdot A}\bigg( \frac{1}{2}\bar{\psi}\cdot \Gamma_{\rho\beta\nu}[\psi,A^{\beta}]\bigg) \nonumber \\
                &=&  Str e^{ik\cdot A} \bigg( +\frac{i}{2}\bar{\psi}\cdot \Gamma_{\nu}[\psi,ik\cdot A]
                     +\frac{1}{2}\bar{\psi}\cdot \ks[\psi,A_{\nu}]\bigg), \nonumber \\
\end{eqnarray}
where we used the equation of motion (\ref{equation_of_motion_for_psi}).  Thus we obtain
\begin{eqnarray}
   k^{\mu}V^h_{\mu\nu}(A,\psi)&=& Str e^{ik\cdot A} \bigg( -i\bar{\psi}\cdot 
                               \Gamma_{\nu}[ik\cdot A,\psi]+2i[F_{\nu\rho},A^{\rho}]\bigg)\nonumber\\
    &=&  Str e^{ik\cdot A} \bigg(+i\{ \psi_{\alpha},\psi_{\beta}\} (\Gamma_0\Gamma_{\nu})_{\alpha\beta}
                    +2i[F_{\nu\rho},A^{\rho}]\bigg)\nonumber \\
       &=&0.      
\end{eqnarray}
where we have used another equation of motion 
\beqa
{[}A^\nu,{[}A_\mu, A_\nu{]}{]}-\frac12(\Gamma_0\Gamma_\mu)_{\alpha\beta}\{ \psi_\alpha,\psi_\beta\}
=0.
\label{equation_of_motion_for_A}
\eeqa

We can similarly show that 
the vertex operator for the antisymmetric tensor field satisfies the conservation law;
multiplying 
$k^{\mu}$, the first term trivially vanishes while the second and third terms cancel
with each other. The fourth term vanishes due to the Jacobi identity. 
Thus the couplings to a graviton and a fourth-rank
antisymmetric tensor $V^h_{\mu\nu}(A,\psi)h^{\mu\nu}(\lambda), V^A_{\mu\nu\rho\sigma}(A,\psi)A^{\mu\nu\rho\sigma}(\lambda)$
are respectively gauge invariant.

\subsection{Vertex operator for gravitino $\Psi^c_{\mu}$}
   The vertex operator for the gravitino $\Psi^c_{\mu}$ can be obtained from the terms with five 
   $\lambda$'s by
using the following relation
\begin{eqnarray}
    V^{\Psi^c}_{\mu}(A,\psi){\Psi^c}^{\mu}(\lambda)&=&
              \frac{1}{5}\left( i(\bar{\lambda}\Gamma_{\mu}\psi)\frac{\delta}{\delta A_{\mu}}
                       -\frac{i}{2}F_{\mu\nu}\Gamma^{\mu\nu}\lambda\frac{\delta}{\delta\psi}\right)\nonumber \\
   &&\phantom{yyyyyy}\bigg( V^h_{\mu\nu}(A,\psi)h^{\mu\nu}(\lambda)
                           +V^A_{\mu\nu\rho\sigma}(A,\psi)A^{\mu\nu\rho\sigma}(\lambda)\bigg).\ \ \
\end{eqnarray}
Many identities are needed in order to derive the vertex operator. Carrying out the complicated calculation, we finally obtain 
\begin{eqnarray}
V^{\Psi^{c}}_{\mu}&=&Str e^{ik\cdot A}\bigg( -\frac{i}{2\cdot 5 !}k^{\lambda}k^{\tau}(\bar{\psi}\cdot{\Gamma_{\mu \lambda}}^{\sigma}\psi)\cdot
(\bar{\psi}\cdot \Gamma_{\nu\tau\sigma}\psi)\cdot \bar{\psi}\Gamma^{\nu}\nonumber \\
 &&+\frac{1}{24}k^{\lambda}(\bar{\psi}\cdot \Gamma_{\lambda\mu\nu}\psi)\cdot 
                                 \bar{\psi}\Gamma^{\nu}\Gamma_{\rho\sigma}\cdot F^{\rho\sigma}
          -\frac{1}{6}k^{\lambda}(\bar{\psi}\cdot \Gamma_{\lambda\alpha\beta}\psi)\cdot 
                                     \bar{\psi}\Gamma^{\beta}\cdot {F^{\alpha}}_{\mu}\nonumber \\
&&+\frac{i}{3}(\bar{\psi}\cdot \Gamma_{\mu}[A_{\nu},\psi])\cdot \bar{\psi}\Gamma^{\nu}
   +\frac{i}{3}(\bar{\psi}\cdot \Gamma_{\nu}[A_{\mu},\psi])\cdot \bar{\psi}\Gamma^{\nu}
   +\frac{i}{6}(\bar{\psi}\cdot \Gamma_{\alpha\beta\mu}\psi)\cdot [A^{\alpha},\bar{\psi}]\Gamma^{\beta}\nonumber \\
 &&-iF_{\mu\nu}\cdot F_{\rho\sigma}\cdot \bar{\psi}\Gamma^{\nu}\Gamma^{\rho\sigma}\bigg).
 \label{VPsi^c}
\end{eqnarray}
It can be shown that this gravitino vertex operator satisfies
\begin{equation}
    k^{\mu}V^{\Psi^c}(A,\psi)=0.
\label{invariance5}
\end{equation}
This assures the gauge invariance of the coupling with gravitino field $V^{\Psi^c}_{\mu}(A,\psi)\Psi^{c\mu}(\lambda)$.
Using the following identity
\begin{equation}
    (\bar{\psi}\Gamma_{\mu}[A_{\nu},\psi])\bar{\psi}\Gamma^{\nu}\Psi^{c\mu}(\lambda)
  =\frac{1}{2}(\bar{\psi}\Gamma_{\alpha}[A_{\mu},\psi])\bar{\psi}\Gamma^{\alpha}\Psi^{c\mu}(\lambda)
         +\frac{1}{4}(\bar{\psi}\Gamma^{\nu\alpha\mu}\psi)[A_{\nu},\bar{\psi}]\Gamma_{\alpha}\Psi^c_{\mu}(\lambda),
\end{equation}
we can rewritten the vertex operator 
\begin{eqnarray}
V^{\Psi^{c}}_{\mu}&=&
     Str e^{ik\cdot A}\bigg( -\frac{i}{2\cdot 5 !}k^{\lambda}k^{\tau}(\bar{\psi}\cdot{\Gamma_{\mu \lambda}}^{\sigma}\psi)\cdot
(\bar{\psi}\cdot \Gamma_{\nu\tau\sigma}\psi)\cdot \bar{\psi}\Gamma^{\nu}\nonumber \\
 &&+\frac{1}{24}k^{\lambda}(\bar{\psi}\cdot \Gamma_{\lambda\mu\nu}\psi)\cdot 
                                 \bar{\psi}\Gamma^{\nu}\Gamma_{\rho\sigma}\cdot F^{\rho\sigma}
          -\frac{1}{6}k^{\lambda}(\bar{\psi}\cdot \Gamma_{\lambda\alpha\beta}\psi)\cdot 
                                     \bar{\psi}\Gamma^{\beta}\cdot {F^{\alpha}}_{\mu}\nonumber \\
     &&+i\bar{\psi}\Gamma_{\mu}[A_{\nu},\psi]\bar{\psi}\Gamma^{\nu}
                 -iF_{\mu\nu}\cdot F_{\rho\sigma}\cdot \bar{\psi}\Gamma^{\nu}\Gamma^{\rho\sigma}\bigg).
\label{VPsi^c2}
\end{eqnarray}

Since the vertex operator for gravitino is the supercurrent, we obtain the supercurrent for I$\hspace{-.1em}$IB
matrix model
\begin{eqnarray}
    J^{(2)}_{\mu}(A,\psi)&=&\mbox{(\ref{VPsi^c}) (or (\ref{VPsi^c2}))}.
\end{eqnarray}
This is conserved due to (\ref{invariance5}).

\subsection{Vertex operator for antisymmetric tensor field $B^c_{\mu\nu}$}
     The sixth order terms give the antisymmetric tensor field $B^c_{\mu\nu}$ vertex operator. 
     After lengthy calculations we finally obtained
the following new expression
\begin{eqnarray}
   V^{B^{c}}_{\mu\nu}(A,\psi)&=& Str e^{ik\cdot A}\bigg( -\frac{1}{8 \cdot 6!}k^{\lambda}k^{\tau}k^{\alpha}(\bar{\psi}
 \cdot {\Gamma_{\mu\lambda}}^{\sigma}\psi) \cdot (\bar{\psi}\cdot \Gamma_{\gamma\tau\sigma} \psi)\cdot 
                        (\bar{\psi}\cdot {\Gamma^{\gamma}}_{\alpha\nu} \psi) \nonumber \\
&& +\frac{i}{64}(\bar{\psi}\cdot\ks \Gamma_{\mu\alpha}\psi)\cdot 
                                                    F^{\alpha\beta}(\bar{\psi}\cdot \ks\Gamma_{\beta\nu}\psi)
+\frac{i}{16\cdot 4!}(\bar{\psi}\cdot\ks \Gamma_{[\mu\alpha}\psi)\cdot(\bar{\psi}\cdot\ks \Gamma^{\alpha\sigma}\psi)
                                                                               \cdot        F_{\sigma\nu]}\nonumber \\
&&-\frac{1}{32}\bar{\psi}\cdot\Gamma_{[\mu}[A^{\sigma},\psi]\cdot (\bar{\psi}\cdot\ks\Gamma_{\sigma\nu]}\psi)
-\frac{1}{64}(\bar{\psi}\cdot\ks\Gamma_{[\mu\alpha}\psi)\cdot \bar{\psi}\Gamma^{\alpha}[A_{\nu]},\psi] \nonumber \\
&&+\frac{i}{4!\cdot 32}\Xi_{\mu\nu\alpha\beta\gamma}\cdot (\bar{\psi}\cdot \Gamma^{\alpha\beta\gamma}\psi)
-\frac{i}{64}[A_{\alpha},F^{\alpha\tau}]\cdot (\bar{\psi}\cdot \Gamma_{\tau\mu\nu}\psi) \nonumber \\
&& +\frac{1}{64}(\bar{\psi}\cdot \Gamma_{\mu\nu\rho\sigma\lambda\tau}\ks \psi)\cdot F^{\rho\sigma}\cdot 
F^{\lambda\tau}+\frac{1}{16}(\bar{\psi} \cdot \Gamma_{\rho\sigma}\ks\psi)\cdot F^{\rho\sigma}\cdot F^{\mu\nu}\nonumber \\
&&-\frac{1}{8}(\bar{\psi}\cdot \Gamma_{\rho\sigma}\ks \psi ) \cdot F^{\mu\rho}\cdot F^{\nu\sigma}
 +\frac{1}{8}(\bar{\psi}\cdot \Gamma_{[\mu\sigma}\ks \psi ) \cdot F^{\sigma\alpha}\cdot 
F_{\alpha\nu]}\nonumber\\
&&-\frac{1}{32}(\bar{\psi}\cdot \Gamma^{\mu\nu}\ks \psi )\cdot F^{\rho\sigma}\cdot F_{\sigma\rho} 
     +\frac{i}{4}\bar{\psi}\cdot \Gamma_{\mu\nu\alpha}[A_{\beta},\psi]\cdot F^{\alpha\beta}\nonumber \\
 && +\frac{i}{8}\bar{\psi}\cdot \Gamma_{\rho\sigma[\mu}[A_{\nu]},\psi]\cdot F^{\rho\sigma} 
   +\frac{i}{4}\bar{\psi}\cdot \Gamma_{(\mu}[A_{\rho)},\psi] \cdot {F^{\rho}}_{\nu} 
                -\frac{i}{4}\bar{\psi}\cdot \Gamma_{(\nu}[A_{\rho)},\psi]\cdot F^{\rho\mu} \nonumber \\
&&-iF_{\mu\rho}\cdot F^{\rho\sigma}\cdot F_{\sigma\nu}+\frac{i}{4}F_{\mu\nu}\cdot F^{\rho\sigma}\cdot F_{\sigma\rho}\bigg),
\end{eqnarray}
where $\Xi_{\mu\nu\rho\sigma\tau}$ is defined by
\begin{equation}
     \Xi_{\mu\nu\rho\sigma\tau}=\{ \psi_{\alpha},\psi_{\beta} \} (\Gamma_0\Gamma_{\mu\nu\rho\sigma\tau})_{\alpha\beta}.
\end{equation}
This vertex operator satisfies
\begin{equation}
    k^{\mu}V^{B^c}_{\mu\nu}(A,\psi)=0.
\end{equation}
Hence the coupling $V^{B^c}_{\mu\nu}(A,\psi)B^{c\mu\nu}(\lambda)$ is gauge invariant.

\subsection{Other vertex operators}
   The next terms with seven $\lambda$'s give the vertex operator for the dilatino $\tilde{\Phi}$.
The calculation becomes more complicated. We have not yet derived the complete expression
so far.
 Here we give a part of our result:
\begin{eqnarray}
V^{\tilde{\Phi}^c}(A,\psi) &=& Str e^{ik\cdot A} \bigg(\frac{1}{8!}(\bar{\psi}\cdot\Gamma^{\alpha\gamma}\ks \psi)
                           \cdot (\bar{\psi}\cdot \Gamma_{\gamma\delta}\ks
\psi)\cdot(\bar{\psi}\cdot \Gamma^{\delta\beta}\ks \psi)\cdot\bar{\psi} \Gamma_{\alpha\beta}\nonumber \\
&&\hskip -3em
 -\frac{i}{2\cdot 5!}F^{\mu\alpha}\cdot (\bar{\psi}\cdot \ks\Gamma_{\alpha\beta}\psi)
                  \cdot (\bar{\psi}\cdot \ks \Gamma^{\beta\nu}\psi)\cdot \bar{\psi}\Gamma_{\mu\nu} 
+\cdot \cdot \cdot \nonumber \\
&&\hskip -3em
  \cdot \cdot \cdot -\frac{1}{8\cdot 4!}F^{\mu\nu}\cdot F^{\rho\sigma}(\bar{\psi}\cdot 
       \Gamma_{\mu\nu\rho\sigma\lambda\alpha\beta}\psi)k^{\lambda}\cdot \bar{\psi}\Gamma^{\alpha\beta} 
-\frac{1}{12}F^{\mu\alpha}\cdot F_{\alpha\beta}\cdot (\bar{\psi}\cdot \ks\Gamma^{\beta\nu}\psi)\cdot 
                                          \bar{\psi}\Gamma_{\mu\nu} \nonumber \\
&&\hskip -3em
-\frac{1}{24}F^{\mu\alpha}\cdot (\bar{\psi}\cdot \ks\Gamma_{\alpha\beta}\psi)\cdot 
                F^{\beta\nu}\cdot \bar{\psi}\Gamma_{\mu\nu} -\frac{1}{48}F^{\rho\sigma}\cdot (\bar{\psi}\cdot \ks 
           \Gamma_{\rho\sigma}\psi)\cdot F^{\mu\nu}\cdot \bar{\psi}\Gamma_{\mu\nu}+\cdot \cdot \cdot \nonumber \\
&&\hskip -3em
 +\cdot \cdot \cdot +\frac{i}{24}\bar{\psi}\cdot \Gamma_{\mu\nu\rho\sigma\lambda\tau}F^{\mu\nu}\cdot
                               F^{\rho\sigma}\cdot F^{\lambda\tau}+i\bar{\psi}\cdot \Gamma^{\mu\nu}
(F_{\mu\rho}\cdot F^{\rho\sigma}\cdot F_{\sigma\nu}-\frac{1}{4}F^{\rho\sigma}\cdot F_{\sigma\rho}\cdot F_{\mu\nu})
           \bigg).\nonumber \\
\end{eqnarray}

The ${\cal O}(\lambda^8)$ terms give the vertex operator 
for the dilaton $\Phi^c$. It has also partly been obtained as follows:
\begin{eqnarray}
V^{\Phi^c}(A,\psi)&=&Str e^{ik\cdot A}\bigg(\frac{1}{8\cdot 8!}(\bar{\psi}\cdot \Gamma^{\alpha\gamma}\ks\psi)
                                                   \cdot (\bar{\psi} \cdot \Gamma_{\gamma\delta}\ks \psi)
\cdot (\bar{\psi}\cdot \Gamma_{\delta\beta}\ks\psi)\cdot (\bar{\psi}\cdot \Gamma_{\alpha\beta}\ks \psi)+\cdot \cdot\cdot \nonumber \\
&&\cdot \cdot \cdot \nonumber \\
&&  +\frac{i}{48}(\bar{\psi}\cdot \Gamma_{\mu\nu\rho\sigma\lambda\tau}\ks\psi ) \cdot F^{\mu\nu}\cdot F^{\rho\sigma}\cdot F^{\lambda\tau} 
 +[A_{\mu},\bar{\psi}]\cdot \Gamma_{\rho\sigma}\Gamma_{\nu}\psi \cdot F^{\mu\nu}\cdot F^{\rho\sigma}\nonumber \\
&&  +\frac{i}{2}(\bar{\psi}\cdot \Gamma^{\mu\nu}\ks \psi)\cdot (F_{\mu\rho}\cdot F^{\rho\sigma}\cdot F_{\sigma\nu}-\frac{1}{4}F^{\rho\sigma}
                                             \cdot F_{\sigma\rho}\cdot F_{\mu\nu})  \nonumber \\
&&  -(F_{\mu\nu}\cdot F^{\nu\rho}\cdot F_{\rho\sigma}\cdot F^{\sigma\mu}-\frac{1}{4}F_{\mu\nu}\cdot F^{\nu\mu}\cdot F_{\rho\sigma} 
\cdot F^{\sigma\rho})\bigg).
\end{eqnarray}

    We summarize our results in Appendix B.

\section{Kinematic factor of four-graviton amplitude: a consistency check} 
As we already mentioned, the I$\hspace{-.1em}$IB matrix model was originally proposed as 
a matrix regularization of the Schild-gauge GS type I$\hspace{-.1em}$IB superstring, and may 
also be regarded as composed of the D-instanton degrees of freedom of typeI$\hspace{-.1em}$IB 
string theory.  We have constructed a set of massless vertex operators of the I$\hspace{-.1em}$IB 
matrix model by requiring the covariance under supersymmetry, and therefore 
they form a massless typeI$\hspace{-.1em}$IB supergravity multiplet by construction.  They also 
appear as factors of one-loop block-block interactions, which serves as a 
consistency check. As a further check on the validity, we will show that the kinematic 
factor of the $R^4$ term may be derived by using our vertex operators, precisely 
in the same manner as was done by using the light-cone supermembrane vertex 
operators\cite{DNP,Plefka}.
 
Suppose that we consider the following four-graviton amplitude
\beqa
\left<
\prod_{r=1}^4 
V^h_{\mu_r\nu_r}(k_r)h^{\mu_r\nu_r}(k_r)
\right>.
\label{Vh4}
\eeqa 
Using the zeromode saturation argument, we find that only the $\psi^4$ term can 
contribute to the amplitude
\beqa
V^h_{\mu\nu}(k)&\sim&-\frac1{4\cdot 4!}k^\lambda k^\tau S
tr e^{ikA}(\bar \psi\cdot \Gamma_{\mu\lambda}^{~~~\!\sigma}\psi)
\cdot (\bar \psi\cdot \Gamma_{\nu\tau\sigma}\psi).
\eeqa
Thus (\ref{Vh4}) contains the factor
\beqa
\int d\psi_{0}^{16} \prod_{r=1}^4 
k_r^{\lambda_r} k_r^{\tau_r} (\bar \psi_0\Gamma_{\mu_r\lambda_r}^{~~~~\sigma_r}\psi_0)
\cdot (\bar \psi_0\Gamma_{\nu_R\tau_r\sigma_r}\psi_0),
\label{t8t8R4}
\eeqa
where $\psi_0$ is the zeromode. 
This is precisely the factor which arises in the four-graviton amplitude in the D-instanton 
background computed in \cite{GreenGutperle}.
Indeed, our construction of the I$\hspace{-.1em}$IB matrix-model vertex operators shares some similarity 
with their construction of the closed string vertices in the D-instanton background.
(\ref{t8t8R4}) is therefore proportional to $t_8t_8R^4$, which is a consistent result. 

\section{Conclusions}
In this paper we have reported progress in determining the complete 
form of massless supergravity vertex operators in the I$\hspace{-.1em}$IB matrix model. 
In principle, they are determined by supersymmetry, but it becomes harder 
to carry out the actual computation for higher vertex operators. 
We have developed two new methods to lighten our work. After 
dozens of pages of Fierz arrangements we have finally reached new formulas 
for the vertex operators emitting two higher component fields in the supergravity
multiplet. 

While it is important to fix the exact form in its own right, vertex operators may be 
useful in computing correlation functions in the I$\hspace{-.1em}$IB matrix model. 
We have taken the same step as in \cite{DNP} to compute a four-point 
graviton amplitude to find a consistent result. 

We have extensively used the fact that the simplest kind of supersymmetric 
Wilson loop operator satisfies the same equation as the generating function of 
the Wilson loop correlation functions does. The one-loop analysis has already revealed that 
the effective action has a form expressed in terms of bilinear of vertex operators.  
In view of these facts, we conjecture that the low-energy effective action of the I$\hspace{-.1em}$IB matrix model 
in the large-$N$ limit is precisely given by tree-level supergravity coupled to the vertex 
operators presented here. It is because they possess the identical ${\cal N}=2$ supersymmetry.


\section*{Acknowledgement}
We thank S.~Iso and H.~Umetsu for useful discussions. 
Y.~K. and S.~M. are supported in part by the Grant-in-Aid for Scientific Research
from the Ministry of Education, Science and Culture of Japan.

\newpage
\section*{Appendix}
\setcounter{section}{0}
\renewcommand{\thesection}{\Alph{section}}
\section{Notation}
 In Appendix A, we present the notation used in this paper. 
 Basically, we follow the same notation as in Ref. \cite{ITU}.

\subsection{10D Gamma matrices}

We use the Minkowskian spacetime metric
\begin{equation}
\eta_{\mu\nu}=diag(-1,+1,\cdot \cdot \cdot ,+1).
\end{equation}
Gamma matrices are defined by 
\begin{equation}
  \{\Gamma_{\mu},\Gamma_{\nu}\} =2 \eta_{\mu\nu}.
\end{equation}
We use the {\it Majorana-Weyl representation}. In this representation $\Gamma_0$ has the following properties:
\begin{eqnarray}
   (\Gamma_{0})^T=\Gamma_0, \\
    (\Gamma_0)^2=-1, \\
    \Gamma_0\Gamma_{\mu}\Gamma_0=-(\Gamma_{\mu})^T.
\end{eqnarray}
The chirality matrix $\Gamma_{11}$ is defined by 
\begin{equation}
\Gamma_{11}\equiv \Gamma_{0} \Gamma_1 \cdot \cdot \cdot \Gamma_{9} ,
\end{equation}
and satisfies
\begin{eqnarray}
  (\Gamma_{11})^2=1, \\
    (\Gamma_{11})^{\dagger}=\Gamma_{11}.
\end{eqnarray}

We denote the anti-symmetrized gamma matrix as follows:
\begin{equation}
\Gamma_{\mu_1\mu_2\cdot \cdot \cdot \mu_n}=\Gamma_{\mu_1}\Gamma_{\mu_2}\cdot \cdot \cdot \Gamma_{\mu_n}
       =\frac{1}{n!}\Gamma_{[\mu_1}\Gamma_{\mu_2}\cdot \cdot \cdot \Gamma_{\mu_n]}.
\end{equation}
\begin{picture}(0,0)
 \put(184,20){\line(1,0){58}}
 \put(184,20){\line(0,1){4}}
 \put(203,20){\line(0,1){4}}
 \put(242,20){\line(0,1){4}}
\end{picture}


\subsubsection{Symmetry and anti-symmetry}
In our representation $\Gamma_0 \Gamma_{\mu_1\cdot \cdot \cdot \mu_n}$ is either symmetric or anti-symmetric. 
For example, 
\begin{eqnarray}
  && \Gamma_0  \cdot \cdot \cdot \mbox{symmetric} \nonumber  \\
  &&\Gamma_0 \Gamma_{\mu} \cdot \cdot \cdot  \mbox{symmetric} \nonumber \\
  && \Gamma_0 \Gamma_{\mu_1\mu_2} \cdot \cdot \cdot  \mbox{anti-symmetric} \nonumber \\
  && \Gamma_0 \Gamma_{\mu_1\mu_2\mu_3} \cdot \cdot \cdot \mbox{anti-symmetric} \nonumber \\
  && \Gamma_0 \Gamma_{\mu_1\mu_2\mu_3\mu_4} \cdot \cdot \cdot \mbox{symmetric} \nonumber \\
  &&\Gamma_0 \Gamma_{\mu_1\mu_2\mu_3\mu_4\mu_5} \cdot \cdot \cdot \mbox{symmetric} .\nonumber 
\end{eqnarray} 

In general,
\begin{equation}
  (\Gamma_0\Gamma_{\mu_1\mu_2\cdot \cdot \cdot \mu_n})_{\alpha\beta}=(-1)^{\frac{n(n-1)}{2}}(\Gamma_0
\Gamma_{\mu_1\mu_2\cdot \cdot \cdot \mu_n})_{\beta\alpha}.
\label{symmetry_anti-symmetry_of_gamma_matrix}
\end{equation}
    
\subsubsection{Duality}
The following relations hold between $\Gamma^{\mu_1\cdot\cdot\cdot \mu_n}$ and $\Gamma^{\mu_{n+1}\cdot \cdot\cdot\mu_{10}}$:
 \begin{equation}
\Gamma^{\mu_1\cdot \cdot \cdot \mu_n}=\frac{(-1)^{\frac{k(k-1)}{2}}}{(10-k)!}\epsilon^{\mu_1\mu_2\cdot \cdot \cdot \mu_{10}}
\Gamma_{\mu_{k+1}\cdot \cdot \cdot \mu_{10}}\Gamma_{11}.
\label{duality_of_gamma_matrix}
\end{equation}


\subsubsection{Multiplication law}
 The product of two anti-symmetrized gamma matrices can be decomposed in terms of anti-symmetrized matrices.
For instance, the product of $\Gamma_{\mu}$ and $\Gamma_{\nu}$ is decomposed as 
\begin{equation}
   \Gamma_{\mu}\Gamma_{\nu}=\Gamma_{\mu\nu}+\eta_{\mu\nu},
\end{equation}
  and the product of $\Gamma_{\mu\nu}$ and $ \Gamma_{\lambda}$ decomposed as
\begin{equation}
  \Gamma_{\mu\nu}\Gamma_{\lambda}=\Gamma_{\mu\nu\lambda}-2\eta_{\mu\lambda}\Gamma_{\nu}.
\end{equation}
\begin{picture}(0,0)(-20,4)
 \put(252,22){\line(1,0){19}}
 \put(252,22){\line(0,1){4}}
 \put(271,22){\line(0,1){4}}
\end{picture}
Generally, the following multiplication law of the gamma matrices holds: 
\begin{equation}
  \Gamma^{\mu_1\mu_2\cdot \cdot \cdot \mu_p}\Gamma^{\nu_1\nu_2\cdot \cdot \cdot \nu_q}=
\sum_{k=0}^{min(p,q)}(-1)^{\frac{1}{2}k(2p-k-1)}\frac{p!q!}{(p-k)!(q-k)!k!}\eta^{\mu_1\nu_1}\!\cdot\!  \cdot \!\cdot 
\eta^{\mu_k\nu_k}\Gamma^{\mu_{k+1}\cdot \cdot \cdot \mu_{p}\nu_{k+1}\cdot \cdot \cdot \nu_{q}}.
\end{equation}
\begin{picture}(0,0)
\put(324,56){\line(1,0){88}}
\put(324,56){\line(0,-1){4}}
\put(360,56){\line(0,-1){4}}
\put(385,56){\line(0,-1){4}}
\put(412,56){\line(0,-1){4}}
\put(332,35){\line(1,0){116}}
\put(332,35){\line(0,1){6}}
\put(368,35){\line(0,1){6}}
\put(434,35){\line(0,1){6}}
\put(448,35){\line(0,1){6}}
\end{picture}
The above two cases correspond to $(p,q)=(1,1)$, $(p,q)=(2,1)$ respectively.


\subsubsection{Commutators and anti-commutators}
  Using the multiplication law, we can derive the following relations:
\begin{equation}
 \{ \Gamma^{\mu},\Gamma^{\nu} \} =2\eta^{\mu\nu} 
\end{equation}
\begin{equation}
 [ \Gamma^{\alpha\beta},\Gamma^{\mu} ]=-4\eta^{\mu\alpha}\Gamma^{\beta} 
\end{equation}
\begin{equation}
 \{ \Gamma^{\alpha\beta\gamma},\Gamma^{\mu} \} =6\eta^{\mu\alpha}\Gamma^{\beta\gamma} 
\end{equation}
\begin{equation}
[\Gamma^{\alpha\beta\gamma\delta},\Gamma^{\mu}]=-8\eta^{\mu\alpha}\Gamma^{\beta\gamma\delta} 
\end{equation}
\begin{equation}
\{\Gamma^{\alpha\beta\gamma\delta\lambda},\Gamma^{\mu} \}=10\eta^{\mu\alpha}\Gamma^{\beta\gamma\delta\lambda}
\end{equation}
\begin{picture}(0,0)(-20,4)
\put(250,114){\line(1,0){15}}
\put(250,114){\line(0,-1){4}}
\put(265,114){\line(0,-1){4}}
\put(248,90){\line(1,0){20}}
\put(248,90){\line(0,-1){4}}
\put(262,90){\line(0,-1){4}}
\put(268,90){\line(0,-1){4}}
\put(250,66){\line(1,0){25}}
\put(250,66){\line(0,-1){4}}
\put(265,66){\line(0,-1){4}}
\put(270,66){\line(0,-1){4}}
\put(275,66){\line(0,-1){4}}
\put(250,42){\line(1,0){28}}
\put(250,42){\line(0,-1){4}}
\put(264,42){\line(0,-1){4}}
\put(269,42){\line(0,-1){4}}
\put(274,42){\line(0,-1){4}}
\put(278,42){\line(0,-1){4}}
\end{picture}%
\begin{equation}
[\Gamma_{\mu\nu},\Gamma_{\alpha\beta}]=-8\eta_{\mu\alpha}\Gamma_{\nu\beta}
\end{equation}
\begin{equation}
[\Gamma_{\mu\nu\rho},\Gamma_{\alpha\beta}]=12\eta_{\mu\alpha}\Gamma_{\nu\rho\beta}
\end{equation}
\begin{equation}
[\Gamma_{\mu\nu\rho\sigma},\Gamma_{\alpha\beta}]=-16\eta_{\mu\alpha}\Gamma_{\nu\rho\sigma\beta}
\end{equation}
\begin{equation}
[\Gamma_{\mu\nu\rho\sigma\gamma},\Gamma_{\alpha\beta}]=20\eta_{\mu\alpha}\Gamma_{\nu\rho\sigma\beta}
\end{equation}
\begin{picture}(0,0)(-20,4)
 \put(245,112){\line(1,0){18}}
\put(245,112){\line(0,-1){6}}
\put(263,112){\line(0,-1){6}}
\put(250,91){\line(1,0){18}}
\put(250,91){\line(0,1){4}}
\put(268,91){\line(0,1){4}}
\put(243,88){\line(1,0){24}}
\put(267,88){\line(0,-1){6}}
\put(243,88){\line(0,-1){6}}
\put(262,88){\line(0,-1){6}}
\put(248,68){\line(1,0){23}}
\put(248,68){\line(0,1){4}}
\put(271,68){\line(0,1){4}}
\put(248,62){\line(1,0){28}}
\put(248,62){\line(0,-1){6}}
\put(267,62){\line(0,-1){6}}
\put(272,62){\line(0,-1){6}}
\put(276,62){\line(0,-1){6}}
\put(253,44){\line(1,0){28}}
\put(253,44){\line(0,1){4}}
\put(281,44){\line(0,1){4}}
\put(246,38){\line(1,0){28}}
\put(246,38){\line(0,-1){6}}
\put(265,38){\line(0,-1){6}}
\put(269.5,38){\line(0,-1){6}}
\put(274,38){\line(0,-1){6}}
\put(251,19){\line(1,0){28}}
\put(251,19){\line(0,1){4}}
\put(279,19){\line(0,1){4}}
\end{picture}
\begin{equation}
\{\Gamma_{\mu\nu\rho},\Gamma_{\alpha\beta\gamma}\}=
18\eta_{\mu\alpha}\Gamma_{\nu\rho\beta\gamma}
-12\eta_{\mu\alpha}\eta_{\nu\beta}\eta_{\rho\gamma}.
\end{equation}
\begin{picture}(0,0)(-20,0)
\put(213,36){\line(1,0){29}}
\put(213,36){\line(0,-1){6}}
\put(237,36){\line(0,-1){6}}
\put(242,36){\line(0,-1){6}}
\put(284,36){\line(1,0){32}}
\put(284,36){\line(0,-1){6}}
\put(300,36){\line(0,-1){6}}
\put(316,36){\line(0,-1){6}}
\put(207,17){\line(1,0){23}}
\put(207,17){\line(0,1){4}}
\put(225,17){\line(0,1){4}}
\put(230,17){\line(0,1){4}}
\put(279,17){\line(1,0){33}}
\put(279,17){\line(0,1){4}}
\put(295,17){\line(0,1){4}}
\put(312,17){\line(0,1){4}}
\end{picture}

\subsubsection{Contractions}
  The following relations among the gamma matrices hold:
\begin{eqnarray}
&& \Gamma^{\alpha}\Gamma_{\alpha}=10\\
&& \Gamma^{\alpha}\Gamma_{\mu}\Gamma_{\alpha}=-8\Gamma_{\mu}\\
&& \Gamma^{\alpha}\Gamma_{\mu\nu}\Gamma_{\alpha}=6\Gamma_{\mu\nu}\\
&& \Gamma^{\alpha}\Gamma_{\mu\nu\rho}\Gamma_{\alpha}=-4\Gamma_{\mu\nu\rho}\\
&& \Gamma^{\alpha}\Gamma_{\mu\nu\rho\sigma}\Gamma_{\alpha}=2\Gamma_{\mu\nu\rho\sigma}\\
&& \Gamma^{\alpha}\Gamma_{\mu\nu\rho\sigma\lambda}\Gamma_{\alpha}=0  \\
&& \Gamma^{\alpha}\Gamma_{\mu\nu\rho\sigma\lambda\tau}\Gamma_{\alpha} =-2\Gamma_{\mu\nu\rho\sigma\lambda\tau} \\
&& \Gamma^{\alpha}\Gamma_{\mu\nu\rho\sigma\lambda\tau\eta}\Gamma_{\alpha}=4\Gamma_{\mu\nu\rho\sigma\lambda\tau\eta}
\end{eqnarray}

\begin{eqnarray}
&& \Gamma^{\alpha\beta}\Gamma_{\alpha\beta}=-90\\
&& \Gamma^{\alpha\beta}\Gamma_{\mu}\Gamma_{\alpha\beta}=-54\Gamma_{\mu}\\
&& \Gamma^{\alpha\beta}\Gamma_{\mu\nu}\Gamma_{\alpha\beta}=-26\Gamma_{\nu}\\
&& \Gamma^{\alpha\beta}\Gamma_{\mu\nu\rho}\Gamma_{\alpha\beta}=-6\Gamma_{\mu\nu\rho}\\
&& \Gamma^{\alpha\beta}\Gamma_{\mu\nu\rho\sigma}\Gamma_{\alpha\beta}=6\Gamma_{\mu\nu\rho\sigma}\\
&& \Gamma^{\alpha\beta}\Gamma_{\mu\nu\rho\sigma\lambda}\Gamma_{\alpha\beta}=10\Gamma_{\mu\nu\rho\sigma\lambda}
\end{eqnarray}

\begin{eqnarray}
&& \Gamma^{\alpha\beta\gamma}\Gamma_{\alpha\beta\gamma}=-720\\
&& \Gamma^{\alpha\beta\gamma}\Gamma_{\mu}\Gamma_{\alpha\beta\gamma}=288\Gamma_{\mu} \\
&& \Gamma^{\alpha\beta\gamma}\Gamma_{\mu\nu}\Gamma_{\alpha\beta\gamma}=-48\Gamma_{\mu\nu} \\
&& \Gamma^{\alpha\beta\gamma}\Gamma_{\mu\nu\rho}\Gamma_{\alpha\beta\gamma}=-48\Gamma_{\mu\nu\rho} \\
&& \Gamma^{\alpha\beta\gamma}\Gamma_{\mu\nu\rho\sigma}\Gamma_{\alpha\beta\gamma}=48\Gamma_{\mu\nu\rho\sigma} \\
&& \Gamma^{\alpha\beta\gamma}\Gamma_{\mu\nu\rho\sigma\lambda}\Gamma_{\alpha\beta\gamma}=0.
\end{eqnarray}

\subsection{Majorana-Weyl spinors and fermion bilinears}
  In ten dimensions, a spinor $\psi$ can be both Weyl and Majorana. {\it Weyl}  means 
that $\psi$ is an eigenstate of the  chirality operator $\Gamma_{11}$,
\begin{equation}
   \Gamma_{11}\psi= \pm \psi,
\end{equation} 
while {\it Majorana} means that the charge conjugate of $\psi$ is $\psi$ itself:
\begin{equation}
  \psi^c =\psi.
\end{equation}
In our representation of gamma matrices we have 
\begin{equation}
\psi^{c}=\psi^{*}.
\end{equation}
Hence the Majorana condition becomes 
\begin{equation}
 \psi^{*}=\psi.
\end{equation}

Let us consider a fermion bilinear $\bar{\psi}_1 \Gamma_{\mu_1\cdot \cdot \cdot \mu_n}\psi_2$.
If $\psi_1$ and $\psi_2$ are Weyl spinors with positive chirality, they satisfy the following relations:
 \begin{eqnarray}
  \bar{\psi}_1 \Gamma_{\mu_1\cdot \cdot \cdot \mu_n}\psi_2 
      &=&\psi_1^{\dagger}\Gamma_0 \Gamma_{\mu_1\cdot \cdot \cdot \mu_n}\Gamma_{11}\psi_2 \nonumber \\
       &=& (-1)^{n+1}\psi^{\dagger}_1\Gamma_{11}\Gamma_0\Gamma_{\mu_1\cdot \cdot \cdot \mu_n}\psi_2 \nonumber \\
       &=& (-1)^{n+1}\bar{\psi}_1\Gamma_{\mu_1\cdot \cdot \cdot \mu_n}\psi_2.
  \end{eqnarray}
Hence, bilinear forms vanish unless $n$ is odd.

Next, let $\psi_1$ and $\psi_2$ be Majorana spinors, then
\begin{eqnarray}
     \bar{\psi}_1\Gamma_{\mu_1\cdot\cdot \cdot \mu_n}\psi_2 
&=& (\psi_1)_{\alpha}(\Gamma_0\Gamma_{\mu_1\cdot \cdot \cdot \mu_n})_{\alpha\beta}(\psi_2)_{\beta}\nonumber \\
&=& -(-1)^{\frac{n(n-1)}{2}}(\psi_2)_{\beta}(\Gamma_0\Gamma_{\mu_1\cdot \cdot \cdot \mu_n})_{\beta\alpha}(\psi_1)_{\alpha}
\nonumber \\
&=& -(-1)^{\frac{n(n-1)}{2}}\bar{\psi}_2\Gamma_{\mu_1\cdot\cdot \cdot \mu_n}\psi_1 .
\end{eqnarray}
If $\psi_1=\psi_2$, the bilinear form vanishes unless $ n=2,3,6,7,10$.

In summary, If $\psi$ is a Majorana-Weyl spinor, the bilinear form 
$\bar{\psi}\Gamma_{\mu_1\cdot\cdot \cdot \mu_n}\psi$ vanishes 
unless $n=3,7$.
      
\subsection{The Fierz identity}
The Fierz identity is given by \cite{Bergshoeff:1981um}
\begin{equation}
 (\bar{\psi}M\chi)(\bar{\lambda}N\psi)=-\frac{1}{32}\sum_{n=0}^{5}C_{n}(\bar{\psi}\Gamma_{\mu_1\cdot \cdot \cdot \mu_n}\phi)
(\bar{\lambda}N\Gamma^{\mu_1\cdot \cdot \cdot \mu_n}M\chi),
\end{equation}
\begin{equation}
C_0=2,\ C_1=2,\ C_2=-1,\ C_3=-\frac{1}{3},C_4=\frac{1}{12},\ C_5=\frac{1}{120}.
\end{equation}

We can derive many identities using the above Fierz rearrangement formula; we present some of them. 
In the following formulas $\lambda$ is a  Majorana-Weyl spinor and $k^2=0$. 
$f^{\mu_1 \cdot \cdot \cdot \mu_n}$ is an anti-symmetric tensor. 

\subsubsection{$\lambda^2$}
The following relations hold:
\begin{eqnarray}
(\bar{\epsilon}\ks \lambda )\bar{\lambda} &=& \frac{1}{96}(\bar{\lambda}\Gamma_{\alpha\beta\gamma}
\lambda )\bar{\epsilon}\ks \Gamma^{\alpha\beta\gamma}\nonumber \\
&=& \frac{1}{16}b_{\alpha\beta}\bar{\epsilon}\Gamma^{\alpha\beta}
-\frac{1}{96}(\bar{\lambda}\Gamma_{\alpha\beta\gamma}\lambda )\bar{\epsilon}
\Gamma^{\alpha\beta\gamma}\ks
\end{eqnarray}
\begin{equation}
(\bar{\epsilon_1}\Gamma^{\rho}\lambda)(\bar{\lambda}\Gamma_{\rho}\epsilon_2)
=-\frac{1}{24}(\bar{\lambda}\Gamma_{\alpha\beta\gamma}\lambda)(\bar{\epsilon_1}\Gamma^{\alpha\beta\gamma}
\epsilon_2)
\end{equation}
\begin{equation}
(\bar{\epsilon}\Gamma^{\nu}\lambda)\bar{\lambda}\ks\Gamma_{\nu}
=\frac{1}{8}b_{\alpha\beta}\bar{\epsilon}\Gamma^{\alpha\beta}
+\frac{1}{48}(\bar{\lambda}\Gamma_{\alpha\beta\gamma}\lambda)\bar{\epsilon}\Gamma^{\alpha\beta\gamma}
\ks
\end{equation}
\begin{equation}
(\bar{\epsilon_1}\ks \lambda)(\bar{\lambda}\ks\epsilon_2)=\frac{1}{16}b^{\alpha\beta}
(\bar{\epsilon_1}\ks \Gamma_{\alpha\beta}\epsilon_2)
\end{equation}
\begin{eqnarray}
 (\bar{\psi} \ks \lambda)^2&=& -\frac{1}{16}\tilde{b}^{\mu\nu}b_{\mu\nu},
\end{eqnarray}
where $b_{\mu\nu}=(\bar{\lambda}\ks \Gamma_{\mu\nu}\lambda)$, and $\tilde{b}_{\mu\nu}=(\bar{\psi}\ks 
\Gamma_{\mu\nu}\psi) $.

\subsubsection{$\lambda^3$}
The following identities hold:
 \begin{equation}
\Gamma^{\alpha\beta}\lambda (\bar{\lambda}\Gamma_{\alpha\beta\gamma}\lambda)=0
\end{equation}
\begin{equation}
\Gamma^{\alpha\beta\gamma}\lambda (\bar{\lambda}\Gamma_{\alpha\beta\gamma}\lambda)=0
\end{equation}


\begin{equation}
    (\bar{\lambda} \Gamma^{\alpha\beta\gamma}\lambda )\bar{\lambda} =\frac{1}{2}(\bar{\lambda}
{\Gamma_{\tau}}^{\alpha\beta}\lambda)\bar{\lambda}\Gamma^{\tau}\Gamma^{\gamma}
\end{equation}
\begin{picture}(0,0)(-20,4)
 \put(240,41){\line(1,0){50}}
 \put(240,41){\line(0,-1){4}}
 \put(247,41){\line(0,-1){4}}
  \put(290,41){\line(0,-1){4}}
  \end{picture}
\begin{equation}
\Gamma_{\mu\nu}\ks\lambda \ b^{\mu\nu}=0
\end{equation}
\begin{equation}
(\bar{\psi}\ks \lambda)^3=-\frac{1}{48}\tilde{b}^{\mu\nu}(\bar{\psi}\Gamma_{\mu}\Gamma_{\alpha}
\ks \lambda) {b^{\alpha}}_{\nu}.
\end{equation}


\subsubsection{$\lambda^4$}
 The following identities hold:
   \begin{equation}
   b_{\mu\nu}b^{\mu\nu}=0
  \end{equation}
 \begin{equation}
     (\bar{\lambda}\Gamma^{\tau\alpha\beta}\lambda){b_{\tau}}^{\gamma}=0
    \end{equation}
   \begin{picture}(0,0)(-20,4)
    \put(198,42){\line(1,0){35}}
     \put(198,42){\line(0,-1){4}}
     \put(205,42){\line(0,-1){4}}
      \put(233,42){\line(0,-1){4}}
    \end{picture}

\begin{eqnarray}
   b_{\mu\nu}b_{\rho\sigma}&=& \frac{1}{3}(b_{\mu\nu}b_{\rho\sigma}+b_{\mu\rho}b_{\sigma\nu}
+b_{\mu\sigma}b_{\nu\rho})\nonumber \\
   && -\frac{2}{3}\eta_{\mu\rho}{b_{\nu}}^{\alpha}b_{\alpha\sigma} \nonumber \\
   && -\frac{1}{3}k_{\mu}{b_{\nu}}^{\alpha}(\bar{\lambda}\Gamma_{\rho\sigma\alpha}\lambda)
\nonumber \\
   && -\frac{1}{3}k_{\rho}{b_{\sigma}}^{\alpha}(\bar{\lambda}\Gamma_{\mu\nu\alpha}\lambda)
\end{eqnarray}
\begin{picture}(0,0)(-20,2)
\put(201,79){\line(1,0){16}}
\put(201,79){\line(0,1){4}}
\put(217,79){\line(0,1){4}}
\put(206,100){\line(1,0){32}}
\put(206,100){\line(0,-1){4}}
\put(238,100){\line(0,-1){4}}
\put(201,52){\line(1,0){12}}
\put(201,52){\line(0,1){4}}
\put(213,52){\line(0,1){4}}
\put(201,24){\line(1,0){12}}
\put(201,24){\line(0,1){4}}
\put(213,24){\line(0,1){4}}
\end{picture}
  \begin{eqnarray}
  b_{\mu\nu}b_{\rho\sigma}&=&b_{\mu\rho}b_{\sigma\nu}\nonumber \\
              && -\eta_{\mu\rho}{b_{\nu}}^{\alpha}b_{\alpha\sigma}\nonumber \\
               && -\frac{1}{2}k_{\mu}{b_{\nu}}^{\alpha}(\bar{\lambda}\Gamma_{\rho\sigma\alpha}
   \lambda)\nonumber \\
             && -\frac{1}{2}k_{\rho}{b_{\sigma}}^{\alpha}(\bar{\lambda}\Gamma_{\mu\nu\alpha}
\lambda)
\end{eqnarray}
\begin{picture}(0,0)(-20,2)
   \put(211,115){\line(1,0){10}}
    \put(211,115){\line(0,-1){4}}
     \put(221,115){\line(0,-1){4}}
    \put(216,74){\line(1,0){15}}
     \put(216,74){\line(0,1){4}}
     \put(231,74){\line(0,1){4}}
     \put(220,95){\line(1,0){33}}
     \put(220,95){\line(0,-1){4}}
     \put(253,95){\line(0,-1){4}}
      \put(225,50){\line(1,0){11}}
      \put(225,50){\line(0,1){4}}
      \put(236,50){\line(0,1){4}}
      \put(225,23){\line(1,0){11}}
      \put(225,23){\line(0,1){4}}
      \put(236,23){\line(0,1){4}}
\end{picture}
\begin{equation}
b^{\nu\tau}(\bar{\lambda}\Gamma_{\nu}\Gamma_{\alpha\beta\gamma}\Gamma_{\tau}\lambda)f^{\alpha\beta\gamma}
=0
\end{equation}

\begin{equation}
b^{\tau\nu}(\bar{\lambda}\Gamma_{\nu}\Gamma_{\mu_1\mu_2\mu_3\mu_4\mu_5}\Gamma_{\tau}\lambda)
f^{\mu_1\mu_2\mu_3\mu_4\mu_5}=40b_{\mu_1\mu_2}(\bar{\lambda}\Gamma_{\mu_3\mu_4\mu_5}\lambda)
f^{\mu_1\mu_2\mu_3\mu_4\mu_5}
\end{equation}
 \begin{eqnarray}
   (\bar{\epsilon}\ks \lambda) \Gamma^{\alpha}\ks \lambda  \ b_{\alpha\mu}
   =\frac{1}{8}\Gamma^{\alpha}\ks \epsilon b_{\mu\beta}{b^{\beta}}_{\alpha}
    +\frac{1}{16}\Gamma^{\alpha\beta\gamma}\ks \epsilon b_{\mu\alpha} b_{\beta\gamma}
  \end{eqnarray}

  \begin{eqnarray}
     (\bar{\psi}\ks \lambda)^4=\frac{1}{4\cdot 96}\tilde{b}^{\mu\alpha}{ {\tilde{b}}_{\alpha}}
    \ ^{\nu} b_{\mu\beta}{b^{\beta}}_{\nu}
      +\frac{1}{8\cdot 96}\tilde{b}^{\mu\nu}\tilde{b}^{\rho\sigma}
     (b_{\mu\nu} b_{\rho\sigma}
   +b_{\mu\rho} b_{\sigma\nu}+ b_{\mu\sigma} b_{\nu\rho})
   \end{eqnarray}

  \begin{equation}
  \frac{1}{5!}\epsilon^{\alpha_1 \alpha_2 \alpha_3 \alpha_4 \alpha_5 \beta_1 \beta_2\beta_3
   \beta_4 \beta_5 }k_{\beta_1}b_{\beta_2 \beta_3}b_{\beta_4\beta_5}=
   k^{\alpha_1}b^{\alpha_2\alpha_3}b^{\alpha_4 \alpha_5}.
 \end{equation}

\subsubsection{$\lambda^5$}
 The following identities hold:
    \begin{eqnarray}
  \Gamma^{\alpha}\Gamma^{\beta}\ks \lambda b_{\mu\alpha}b_{\beta\nu} = 
\frac{3}{5}\ks \Gamma_{\mu}\Gamma^{\alpha}\lambda b_{\alpha\beta}{b^{\beta}}_{\nu}
 -\frac{2}{5}\ks \Gamma_{\nu}\Gamma^{\alpha}\lambda b_{\alpha\beta}{b^{\beta}}_{\mu}
    \end{eqnarray}

    \begin{eqnarray}
    \lambda b_{\mu\alpha}{b^{\alpha}}_{\nu}&=&\frac{1}{10}\Gamma_{(\mu}\Gamma^{\alpha}\lambda 
{b_{\alpha}}^{\beta}b_{\beta\nu)}+\frac{1}{10}\ks \Gamma^{\alpha}\lambda (\bar{\lambda}\Gamma_{\alpha(\mu \beta}
\lambda ){b^{\beta}}_{\nu)}\nonumber \\
&=&\frac{1}{6}\Gamma_{(\mu}\Gamma^{\alpha}\lambda {b_{\alpha}}^{\beta}b_{\beta\nu)}
-\frac{1}{6}\Gamma^{\alpha}\ks \lambda(\bar{\lambda}\Gamma_{\alpha(\mu\beta}\lambda){b^{\beta}}_{\nu)}
    \end{eqnarray}
    \begin{eqnarray}
     (\bar{\epsilon}\ks \lambda) b_{\mu\alpha}{b^{\alpha}}_{\nu}&=&\frac{1}{10}(\bar{\epsilon}\ks \Gamma_{(\mu}
\Gamma^{\beta}\lambda )b_{\beta\alpha}{b^{\alpha}}_{\nu)} \\
&=&\frac{1}{10}(\bar{\epsilon}\Gamma_{(\mu}\Gamma_{\alpha}\ks \lambda ){b^{\alpha}}_{\beta}
{b^{\beta}}_{\nu)}
  +\frac{1}{5}k_{(\mu}(\bar{\epsilon}\Gamma^{\alpha}\lambda)b_{\alpha\beta}{b^{\beta}}_{\nu)}
     \end{eqnarray}

\begin{equation}
\Gamma^{\alpha\beta}\lambda b_{\alpha\nu}(\bar{\lambda}\Gamma_{\rho\sigma\beta}
\lambda )f^{\nu\rho\sigma}=-2\Gamma_{\nu}\Gamma^{\alpha}\lambda (\bar{\lambda}
\Gamma_{\alpha\rho\beta}\lambda ){b^{\beta}}_{\sigma}f^{\nu\rho\sigma}
\end{equation}

\begin{eqnarray}
\Gamma^{\alpha\beta}\lambda b_{\rho\alpha}b_{\beta\sigma} f^{\rho\sigma}
&=& \Gamma_{\rho}\Gamma^{\alpha}\lambda b_{\alpha\beta}{b^{\beta}}_{\sigma} f^{\rho\sigma} 
+\ks \Gamma^{\alpha}\lambda(\bar{\lambda}\Gamma_{\alpha\rho\beta}\lambda){b^{\beta}}_{\sigma}f^{\rho\sigma}
\nonumber \\
&=& \Gamma_{\rho}\Gamma^{\alpha}\lambda b_{\alpha\beta}{b^{\beta}}_{\sigma} f^{\rho\sigma} 
-\Gamma^{\alpha}\ks \lambda (\bar{\lambda}\Gamma_{\alpha\rho\beta}\lambda){b^{\beta}}_{\sigma}f^{\rho\sigma}
\end{eqnarray}


\begin{equation}
\Gamma_{\alpha\beta}\ks\lambda b^{\mu\alpha}b^{\beta\nu}=\ks\Gamma^{\mu}\Gamma^{\alpha}
\lambda b_{\alpha\beta}b^{\beta\nu}
\end{equation}
\begin{picture}(0,0)(-20,0)
\put(235,38){\line(1,0){52}}
\put(235,38){\line(0,-1){4}}
\put(287,38){\line(0,-1){4}}
\end{picture}

\begin{eqnarray}
\Gamma^{\alpha}\lambda b_{\alpha\nu}b_{\rho\sigma}f^{\nu\rho\sigma}
&=& -\frac{1}{5}\Gamma_{\nu}\Gamma^{\alpha\beta}\lambda b_{\rho\alpha}b_{\beta\sigma}
f^{\nu\rho\sigma} 
 -\frac{1}{10}\ks \Gamma^{\alpha}\Gamma^{\beta}\lambda b_{\alpha\nu}(\bar{\lambda}
\Gamma_{\rho\sigma\beta}\lambda)f^{\nu\rho\sigma}\nonumber \\
&=& -\frac{1}{5}\Gamma_{\nu\rho}\Gamma^{\alpha}\lambda b_{\alpha\beta}{b^{\beta}}_{\sigma}
f^{\nu\rho\sigma}
\nonumber \\
&& -\frac{2}{5}k_{\nu}\Gamma^{\alpha}\lambda (\bar{\lambda}\Gamma_{\alpha\rho\beta}
\lambda ){b^{\beta}}_{\sigma}f^{\nu\rho\sigma}\nonumber \\
&&+\frac{2}{5}\ks \Gamma_{\nu}\Gamma^{\alpha}\lambda(\bar{\lambda}\Gamma_{\alpha\rho\beta}\lambda
){b^{\beta}}_{\sigma}f^{\nu\rho\sigma}
\end{eqnarray}

\begin{eqnarray}
(\bar{\epsilon}\ks \lambda )b_{\mu\nu}b_{\rho\sigma}f^{\mu\nu\rho\sigma} &=&
 -\frac{1}{15}(\bar{\epsilon}\ks \Gamma_{\mu\nu\rho}\Gamma^{\alpha}\lambda )
b_{\alpha\beta}{b^{\beta}}_{\sigma}f^{\mu\nu\rho\sigma}\nonumber \\
&& -\frac{2}{5}k_{\mu}(\bar{\epsilon}\Gamma_{\nu}\ks \Gamma_{\alpha}\lambda )(\bar{\lambda}
{\Gamma^{\alpha}}_{\rho\gamma}\lambda ){b^{\gamma}}_{\sigma} f^{\mu\nu\rho\sigma}
\end{eqnarray}


\begin{eqnarray}
\Gamma^{\mu\nu\rho}\Gamma_{\alpha\beta}\lambda b_{\mu\nu}b_{\rho\sigma}f^{\alpha\beta}
&=&
 -\frac{2}{5}\Gamma_{\alpha\beta}\Gamma^{\delta}\lambda b_{\delta\eta}b^{\eta\sigma}f^{\alpha\beta}
\nonumber \\
&&+\frac{24}{5}\eta_{\alpha\sigma}\Gamma^{\delta}\lambda b_{\delta\eta}{b^{\eta}}_{\beta}f^{\alpha\beta}
\nonumber \\
&&-\frac{4}{5}\Gamma_{\sigma}\Gamma_{\alpha}\Gamma^{\delta}\lambda b_{\delta\eta}{b^{\eta}}_{\beta}f^{\alpha\beta}
\nonumber \\
&&-\frac{8}{5}\Gamma_{\alpha}\ks \Gamma^{\delta}\lambda (\bar{\lambda}\Gamma_{\delta\beta\eta}\lambda)
{b^{\eta}}_{\sigma}f^{\alpha\beta}  \nonumber \\
&& -\frac{16}{5}k_{\sigma}\Gamma^{\delta}\lambda (\bar{\lambda}\Gamma_{\delta\alpha\eta}\lambda){b^{\eta}}_{\beta}
f^{\alpha\beta} \nonumber \\
&&-\frac{32}{5}k_{\alpha}\Gamma^{\delta}\lambda (\bar{\lambda}\Gamma_{\delta\beta\eta}\lambda ){b^{\eta}}_{\sigma}
f^{\alpha\beta}\nonumber \\
&& +\frac{16}{5}\Gamma_{\sigma}\ks\Gamma^{\delta}\lambda (\bar{\lambda}\Gamma_{\delta\alpha\eta}\lambda){b^{\eta}}_{\beta}
f^{\alpha\beta}
\end{eqnarray}
\begin{picture}(0,0)(-20,0)
\put(288,107){\line(1,0){33}}
\put(288,107){\line(0,1){4}}
\put(321,107){\line(0,1){4}}
\put(287,51){\line(1,0){33}}
\put(287,51){\line(0,1){4}}
\put(320,51){\line(0,1){4}}
\end{picture}

\begin{eqnarray}
\ks \lambda {b_{\tau}}^{\nu} (\bar{\lambda}\Gamma^{\alpha\beta\tau}\lambda )&=&
 -\frac{1}{10}\ks \Gamma^{\nu}\Gamma_{\delta}\lambda b^{\delta\tau}(\bar{\lambda}
{\Gamma_{\tau}}^{\alpha\beta}\lambda)+\frac{1}{5}\ks \Gamma^{\alpha}\Gamma_{\delta}
\lambda (\bar{\lambda}\Gamma^{\delta\beta\tau}\lambda ){b_{\tau}}^{\nu}
\end{eqnarray}
\begin{picture}(0,0)(-20,0)
\put(305,45){\line(1,0){48}}
\put(305,45){\line(0,-1){4}}
\put(353,45){\line(0,-1){4}}
\end{picture}

\begin{equation}
\Gamma_{\alpha}\Gamma_{\nu}\ks \lambda {b_{\tau}}^{\nu}(\bar{\lambda}\Gamma^{\alpha\beta\tau}
\lambda )=-2\Gamma^{\nu}\lambda {b_{\nu}}^{\tau}{b_{\tau}}^{\beta}
\end{equation}

\begin{equation}
  (\bar{\psi}\ks \lambda)^5=\frac{1}{4\cdot 5\cdot 48}\tilde{b}^{\mu\alpha}\tilde{b}_{\alpha}
\ ^{\nu}(\bar{\psi}\Gamma_{\mu}\Gamma_{\beta}\ks\lambda){b^{\beta}}_{\gamma}{b^{\gamma}}_{\nu}.
\end{equation}

\subsubsection{$\lambda^6$}
 The following identities hold:
\begin{equation}
  b_{\alpha\beta}b^{\beta\gamma}{b_{\gamma}}^{\alpha}=0
\end{equation}

\begin{eqnarray}
b_{\rho\sigma}b_{\mu\alpha}{b^{\alpha}}_{\nu}S^{\mu\nu}f^{\rho\sigma}&=&
-\frac{8}{15}\eta_{\rho\mu}b_{\sigma\beta}{b^{\beta}}_{\alpha}{b^{\alpha}}_{\nu}S^{\mu\nu}f^{\rho\sigma}
\nonumber \\
&& -\frac{1}{15}\eta_{\mu\nu}{b_{\rho}}^{\beta}b_{\beta\alpha}{b^{\alpha}}_{\sigma}S^{\mu\nu}f^{\rho\sigma}
\nonumber \\
&&+\frac{7}{15}k_{\mu}{b_{\nu}}^{\alpha}(\bar{\lambda}\Gamma_{\alpha\rho\beta}\lambda){b^{\beta}}_{\sigma}
S^{\mu\nu}f^{\rho\sigma}\nonumber \\
&&-\frac{2}{15}k_{\mu}{b_{\rho}}^{\alpha}(\bar{\lambda}\Gamma_{\alpha\nu\beta}\lambda){b^{\beta}}_{\sigma}
S^{\mu\nu}f^{\rho\sigma}\nonumber \\
&&+\frac{1}{15}k_{\mu}{b_{\rho}}^{\alpha}(\bar{\lambda}\Gamma_{\alpha\sigma\beta}\lambda){b^{\beta}}_{\nu}
S^{\mu\nu}f^{\rho\sigma}\nonumber \\
&&+\frac{8}{15}k_{\rho}{b_{\mu}}^{\alpha}(\bar{\lambda}\Gamma_{\alpha\nu\beta}\lambda){b^{\beta}}_{\sigma}
S^{\mu\nu}f^{\rho\sigma}
\end{eqnarray}
\begin{eqnarray}
(\bar{\lambda}\ks\epsilon) \Gamma_{\alpha}\ks\lambda b^{\alpha\beta}b_{\beta\mu}
&=& 
\frac{1}{8}k^{\alpha}\Gamma_{\alpha\beta}\epsilon{b^{\beta}}_{\gamma}b^{\gamma\delta}b_{\delta\mu}
\nonumber \\
&&-\frac{1}{48}\Gamma_{\mu\nu\rho\sigma}\epsilon k^{\nu}b^{\rho\alpha}b_{\alpha\beta}b^{\beta\sigma}
\nonumber \\
&&
-\frac{1}{48}k_{\mu}\Gamma^{\nu\rho\sigma\lambda}\epsilon k_{\nu}{b_{\rho}}^{\alpha}
(\bar{\lambda}\Gamma_{\alpha\sigma\beta}\lambda){b^{\beta}}_{\lambda}
\end{eqnarray}
\begin{equation}
(\bar{\lambda}\ks \Gamma^{\mu}\Gamma^{\theta}\epsilon)\Gamma_{\alpha}\ks\lambda b^{\alpha\beta}b_{\beta\mu}
=\frac{1}{4}\Gamma^{\nu\beta\mu}\Gamma^{\theta}\epsilon k_{\nu}{b_{\beta}}^{\gamma}b_{\gamma\delta}{b^{\delta}}_{\mu}
\end{equation}
\begin{eqnarray}
(\bar{\lambda}\Gamma^{\alpha}\epsilon_1)(\bar{\epsilon_2}\Gamma^{\rho}\Gamma_{\beta}\ks \lambda )
{b^{\beta}}_{\delta}b^{\delta\mu}
&=&\frac{1}{96}(\bar{\epsilon_2}\Gamma^{\rho}k^{\lambda}\Gamma_{\beta\lambda\mu_1\mu_2\mu_3}\Gamma^{\alpha}
\epsilon_1)(\bar{\lambda}\Gamma^{\mu_1\mu_2\mu_3}\lambda){b^{\beta}}_{\delta}b^{\delta\mu}
\nonumber \\
&&-\frac{3}{32}(\bar{\epsilon_2}\Gamma^{\rho}k^{\lambda}\Gamma_{\lambda\mu_2\mu_3}\Gamma^{\alpha}\epsilon_1)
{b^{\mu_2}}_{\delta}(\bar{\lambda}\Gamma^{\delta\mu_3\eta}\lambda){b_{\eta}}^{\mu}\nonumber \\
&&+\frac{1}{96}k^{\mu}(\bar{\epsilon_2}\Gamma^{\rho}\Gamma_{\mu_1\mu_2\mu_3}\Gamma^{\alpha}\epsilon_1){b^{\mu_1}}_{\delta}
(\bar{\lambda}\Gamma^{\delta\mu_2\eta}\lambda){b_{\eta}}^{\mu_3}\nonumber \\
&&-\frac{1}{24}(\bar{\epsilon_2}\Gamma^{\rho}k^{\lambda}\Gamma_{\lambda\mu_2\mu_3}\Gamma^{\alpha}\epsilon_1){b^{\mu_2}}_{\delta}
(\bar{\lambda}\Gamma^{\delta\mu\eta}\lambda){b_{\eta}}^{\mu_3}\nonumber \\
&&-\frac{1}{96}(\bar{\epsilon_2}\Gamma^{\rho}{\Gamma^{\mu}}_{\mu_2\mu_3}\Gamma^{\alpha}\epsilon_1)
b^{\mu_2\delta}b_{\delta\eta}b^{\eta\mu_3} \nonumber \\
&&-\frac{1}{16}(\bar{\epsilon_2}\Gamma^{\rho}\Gamma_{\mu_3}\Gamma^{\alpha}\epsilon_1){b^{\mu_3}}_{\beta}{b^{\beta}}_{\delta}b^{\delta\mu}
\end{eqnarray}
\begin{picture}(0,0)(-20,6)
\put(314,157){\line(1,0){73}}
\put(314,157){\line(0,-1){4}}
\put(352,157){\line(0,-1){4}}
\put(387,157){\line(0,-1){4}}
\put(318,128){\line(1,0){73}}
\put(318,128){\line(0,-1){4}}
\put(356,128){\line(0,-1){4}}
\put(391,128){\line(0,-1){4}}
\end{picture}
\begin{eqnarray}
(\bar{\lambda}\Gamma_{\mu}\epsilon_1)(\bar{\epsilon_2}\Gamma^{\rho}\Gamma_{\beta}\ks\lambda){b^{\beta}}_{\delta}
b^{\delta\mu} &=& -\frac{1}{8}(\bar{\epsilon_2}\Gamma^{\rho}k^{\lambda}\Gamma_{\lambda\mu_1\mu_2\mu_3}\epsilon_1)
{b^{\mu_1}}_{\delta}(\bar{\lambda}\Gamma^{\delta\mu_2\eta}\lambda ){b_{\eta}}^{\mu_3} \nonumber \\
&&-\frac{1}{8}(\bar{\epsilon_2}\Gamma^{\rho}\Gamma_{\mu_2\mu_3}\epsilon_1)b^{\mu_2\delta}b_{\delta\eta}b^{\eta\mu_3}
\end{eqnarray}
\begin{eqnarray}
b_{\rho\sigma}{b_{\lambda}}^{\beta}b_{\beta\mu}f^{\rho\sigma\lambda}&=&
-\frac{1}{3}\eta_{\mu\rho}{b_{\sigma}}^{\alpha}b_{\alpha\beta}{b^{\beta}}_{\lambda}
f^{\rho\sigma\lambda}\nonumber \\
&&+\frac{1}{3}k_{\mu}{b_{\rho}}^{\alpha}(\bar{\lambda}\Gamma_{\alpha\sigma\beta}\lambda){b^{\beta}}_{\lambda}
f^{\rho\sigma\lambda}\nonumber \\
&&-\frac{1}{3}k_{\rho}{b_{\sigma}}^{\alpha}(\bar{\lambda}\Gamma_{\alpha\mu\beta}\lambda){b^{\beta}}_{\lambda}
f^{\rho\sigma\lambda}
\end{eqnarray}
\begin{equation}
(\bar{\psi}\ks \lambda)^6=-\frac{1}{8\cdot 6 !}\tilde{b}^{\mu\alpha}\tilde{b}_{\alpha\beta}
\tilde{b}^{\beta\nu}b_{\mu\gamma}b^{\gamma\delta}b_{\delta\nu}
\end{equation}

\begin{equation}
\frac{1}{5!}\epsilon^{\alpha_1\alpha_2\alpha_5\beta_1\beta_2\beta_3\beta_beta_5\alpha_3\alpha_4}
k_{\beta_1}b_{\beta_2\beta_3}b_{\beta_4\beta_5}b_{\alpha_3\alpha_4}=
 k^{\alpha_1}b^{\alpha_2\alpha_3}b_{\alpha_3\alpha_4}b^{\alpha_4\alpha_5},
\end{equation}
where $S^{\mu_1\cdot \cdot \cdot \mu_n}$ is a symmetric tensor.

\subsubsection{$\lambda^7$}
The following identities hold:

  \begin{equation}
   (\bar{\epsilon}\ks \lambda)b_{\mu\rho}b^{\rho\sigma}b_{\sigma\nu}=\frac{1}{3}(\bar{\epsilon}
\Gamma^{\rho}\Gamma_{\alpha}\ks \lambda) b_{\mu\rho}b^{\alpha\sigma}b_{\sigma\nu}
   \end{equation}
   \begin{picture}(0,0)(-20,0)
     \put(273,16){\line(1,0){36}}
      \put(273,16){\line(0,1){4}}
      \put(309,16){\line(0,1){4}}
    \end{picture}

\begin{equation}
 (\bar{\epsilon}\Gamma_{\rho}\Gamma_{\alpha}\ks \lambda) {b^{\alpha}}_{\mu}b^{\rho\sigma}
b_{\sigma\nu}=(\bar{\epsilon}\ks \lambda) b_{\mu\rho}b^{\rho\sigma}b_{\sigma\nu}
\end{equation}

\begin{equation}
(\bar{\epsilon}\ks \lambda)b_{\mu\rho}b^{\rho\sigma}b_{\sigma\nu}f^{\mu\nu}
= \frac{1}{7}(\bar{\epsilon}\ks \Gamma_{\mu}\Gamma_{\alpha}\lambda ){b^{\alpha}}_{\rho}b^{\rho\sigma}b_{\sigma\nu}
f^{\mu\nu}
\end{equation}

\begin{eqnarray}
\Gamma^{\mu}\lambda b_{\mu\alpha}b^{\alpha\beta}b_{\beta\nu}&=&-\frac{1}{8}\Gamma_{\nu}\Gamma^{\alpha\beta}\lambda 
b_{\alpha\gamma}b^{\gamma\delta}b_{\delta\beta}
-\frac{1}{8}\ks \Gamma^{\alpha\beta}\lambda b_{\alpha\gamma}b^{\gamma\delta}(\bar{\lambda}\Gamma_{\delta\nu\beta}\lambda)
\nonumber \\
&=&-\frac{1}{8}\Gamma_{\nu}\Gamma^{\alpha\beta}\lambda 
b_{\alpha\gamma}b^{\gamma\delta}b_{\delta\beta}-\frac{1}{8}\ks \Gamma^{\alpha\beta}\lambda {b_{\alpha}}^{\gamma}
(\bar{\lambda}\Gamma_{\gamma\nu\delta}\lambda){b^{\delta}}_{\beta}
\end{eqnarray}

\begin{eqnarray}
\Gamma_{\mu}\ks\lambda b^{\mu\rho}b_{\rho\sigma}b^{\sigma\nu}& =&     
\frac{1}{8}\ks \Gamma_{\nu}\Gamma^{\alpha\beta}\lambda b_{\alpha\gamma}b^{\gamma\delta}
b_{\delta\beta}          
\nonumber \\
&=& -\frac{1}{8}\Gamma^{\nu}\Gamma_{\alpha\beta}\ks \lambda {b^{\alpha}}_{\rho}
b^{\rho\sigma}{b_{\sigma}}^{\beta}+\frac{1}{4}k_{\nu}\Gamma_{\alpha\beta}\lambda {b^{\alpha}}_{\rho}
b^{\rho\sigma}{b_{\sigma}}^{\beta}
\end{eqnarray}

\begin{eqnarray}
(\bar{\epsilon}\ks \lambda) b_{\mu\rho}b^{\rho\sigma}b_{\sigma\nu} f^{\mu\nu} &=& -\frac{1}{56}(\bar{\epsilon}
\Gamma_{\mu\nu}\Gamma_{\alpha\beta}\ks \lambda){b^{\alpha}}_{\rho}b^{\rho\sigma}{b_{\sigma}}^{\beta}f^{\mu\nu}
\nonumber \\
&& -\frac{1}{28}k_{\mu}(\bar{\epsilon}\Gamma_{\nu}\Gamma_{\alpha\beta}\lambda){b^{\alpha}}_{\rho}b^{\rho\sigma}
{b_{\sigma}}^{\beta} f^{\mu\nu}\nonumber \\
&&+\frac{2}{7}k_{\mu}(\bar{\epsilon}\Gamma_{\alpha}\lambda){b^{\alpha}}_{\rho}b^{\rho\sigma}b_{\sigma\nu}f^{\mu\nu}
\nonumber \\
&=&-\frac{1}{56}(\bar{\epsilon}
\Gamma_{\mu\nu}\Gamma_{\alpha\beta}\ks \lambda){b^{\alpha}}_{\rho}b^{\rho\sigma}{b_{\sigma}}^{\beta}f^{\mu\nu}
\nonumber \\
&&-\frac{1}{14}k_{\mu}(\bar{\epsilon}\Gamma_{\nu}\Gamma_{\alpha\beta}\lambda){b^{\alpha}}_{\rho}b^{\rho\sigma}
{b_{\sigma}}^{\beta} f^{\mu\nu} \nonumber \\
&&-\frac{1}{28}k_{\mu}(\bar{\epsilon}\ks \Gamma^{\alpha\beta}\lambda){b_{\alpha}}^{\gamma}(\bar{\lambda}
\Gamma_{\gamma\nu\delta}\lambda ){b^{\delta}}_{\beta}f^{\mu\nu}
\end{eqnarray}

\begin{eqnarray}
\Gamma_{\alpha\delta}\lambda {b^{\delta}}_{\mu}b^{\alpha\beta}b_{\beta\nu}f^{\mu\nu} &=&
\frac{1}{3}\Gamma_{\mu\delta}\lambda b^{\delta\theta}b_{\theta\beta}{b^{\beta}}_{\nu}f^{\mu\nu}\nonumber \\
&&-\frac{1}{3}k^{\delta}\Gamma_{\alpha\delta}\lambda {b_{\mu}}^{\theta}(\bar{\lambda}
\Gamma_{\theta\alpha\beta}\lambda){b^{\beta}}_{\nu}f^{\mu\nu} \nonumber \\
&&+\frac{1}{3}k^{\alpha}\Gamma_{\alpha\delta}\lambda(\bar{\lambda}{\Gamma^{\delta}}_{\mu\theta}\lambda ){b^{\theta}}_{\beta}
{b^{\beta}}_{\nu}f^{\mu\nu}
\end{eqnarray}

\begin{eqnarray}
\Gamma^{\delta}\ks \lambda (\bar{\lambda}\Gamma_{\delta\mu\alpha}\lambda )b^{\alpha\beta}b_{\beta\nu}f^{\mu\nu}
&=& \frac{1}{20}\ks \Gamma_{\nu}\Gamma_{\alpha\beta}\lambda b^{\alpha\delta}(\bar{\lambda}\Gamma_{\delta\mu\tau}\lambda)b^{\tau\beta}
f^{\mu\nu} \nonumber \\
&&-\frac{11}{5}\lambda b_{\mu\delta}{b^{\delta}}_{\tau}{b^{\tau}}_{\nu} f^{\mu\nu}\nonumber \\
&& +\frac{3}{5}\Gamma_{\mu}\Gamma^{\eta}\lambda b_{\eta\tau}b^{\tau\delta}b_{\delta\nu}f^{\mu\nu}
\end{eqnarray}

\begin{eqnarray}
\Gamma_{\alpha}\ks \lambda b^{\alpha\delta}(\bar{\lambda}\Gamma_{\delta\mu\tau}\lambda ){b^{\tau}}_{\nu}f^{\mu\nu}
&=& \frac{3}{20}\ks \Gamma_{\nu}\Gamma_{\alpha\beta}\lambda b^{\alpha\delta}(\bar{\lambda}\Gamma_{\delta\mu\tau}\lambda )
b^{\tau\beta}f^{\mu\nu} \nonumber \\
&&+\frac{7}{5}\lambda b_{\mu\delta}{b^{\delta}}_{\tau}{b^{\tau}}_{\nu}f^{\mu\nu} \nonumber \\
&&-\frac{1}{5}\Gamma_{\mu}\Gamma^{\eta}\lambda b_{\eta\tau}b^{\tau\delta}b_{\delta\nu}f^{\mu\nu}
\end{eqnarray}

\begin{eqnarray}
\Gamma_{\tau\eta}\lambda {b^{\eta}}_{\mu}b^{\tau\delta}b_{\delta\nu}f^{\mu\nu}&=&
-\frac{1}{20}\ks \Gamma_{\nu}\Gamma_{\alpha\beta}\lambda b^{\alpha\delta}(\bar{\lambda}\Gamma_{\delta\mu\tau}\lambda)
b^{\tau\beta}f^{\mu\nu} \nonumber \\
&&-\frac{4}{5}\lambda b_{\mu\delta}{b^{\delta}}_{\tau}{b^{\tau}}_{\nu}f^{\mu\nu}\nonumber \\
&&+\frac{2}{5}\Gamma_{\mu}\Gamma^{\eta}\lambda b_{\eta\tau}b^{\tau\delta}b_{\delta\nu}f^{\mu\nu}
\end{eqnarray}

  \begin{equation}
      (\bar{\psi} \ks \lambda)^7 =\frac{1}{8\cdot 8!}\tilde{b}^{\alpha\gamma}
\tilde{b}_{\gamma\delta}\tilde{b}^{\delta\beta}(\bar{\psi}\Gamma_{\alpha\beta}\Gamma_{\mu\nu}\ks 
\lambda )b^{\mu\rho}b_{\rho\sigma}b^{\sigma\nu}.
   \end{equation}

\subsubsection{$\lambda^8$}
The following identities hold:
 \begin{equation}
    b_{\alpha\beta}b^{\beta\gamma}b_{\gamma\delta}b^{\delta\alpha}=-b_{\mu\nu}b^{\mu\rho}b_{\rho\sigma}
b^{\sigma\nu}
   \end{equation}

\begin{equation}
    \Gamma^{\alpha\mu\nu\beta\gamma}\epsilon k_{\alpha}b_{\mu\nu}b_{\beta\rho}b^{\rho\sigma}
b_{\sigma\gamma}=0
\end{equation}

\begin{equation}
    (\bar{\epsilon} \ks \lambda)\Gamma^{\mu\nu}\ks \lambda b_{\mu\rho}b^{\rho\sigma}b_{\sigma\nu}
   =-\frac{1}{8}\ks \epsilon b_{\mu\rho}b^{\rho\sigma}b_{\sigma\nu}b^{\nu\mu}
\end{equation}

\begin{equation}
   (\bar{\psi} \ks \lambda)^8 =-\frac{1}{8^2\cdot 8!} \tilde{b}^{\alpha\gamma}\tilde{b}_{\gamma\delta}
  \tilde{b}^{\delta\beta}\tilde{b}_{\alpha\beta}b_{\mu\rho}b^{\rho\sigma}b_{\sigma\nu}b^{\nu\mu}.
\end{equation}

\subsection{The symmetrized trace}
The symmetrized trace "$Str$" is defined by 
\begin{eqnarray}
  Str e^{ik\cdot A}B_1\cdot B_2 \cdot \cdot \cdot B_n &=& \int_{0}^{1}dt_1\int_{t_1}^{1}dt_2\cdot \cdot \cdot \int_{t_{n-2}}^1 dt_{n-1}
  \nonumber \\ &&\times tr e^{ik\cdot A t_1}B_1e^{ik\cdot A(t_2-t_1)}B_2
\cdot \cdot \cdot e^{ik\cdot A (t_{n-1}-t_{n-2})}B_{n-1}e^{ik\cdot A(1-t_{n-1})}B_n \nonumber \\
&&+(permutations \  of \ B_{i} \ 's(i=2,3\cdot \cdot \cdot n)).
\end{eqnarray}
The dot on the left hand side indicates that the operators $B_i$ are symmetrized.
If we set $k=0$, the symmetrized trace becomes
\begin{equation}
   Str (B_1\cdot B_2\cdot \cdot \cdot \cdot B_n)=\frac{1}{n!}\sum_{perm.}tr(B_{i_1}B_{i_2}\cdot \cdot \cdot B_{i_n}).
\end{equation}
The explicit forms with two and three inserted operators are written as 
\begin{equation}
Str(e^{ik\cdot A}B\cdot C)=tr\int_0^1dt e^{ik\cdot At}Be^{ik\cdot A(1-t)}C,
\end{equation}
\begin{eqnarray}
Str (e^{ik\cdot A}B\cdot C\cdot D)&=& tr\int_0^1dt_1\int_{t_1}^1dt_2e^{ik\cdot A t_1}Be^{ik\cdot A(t_2-t_1)}Ce^{ik\cdot A(1-t_2)}D
    \nonumber \\  && +(C\leftrightarrow D),
\end{eqnarray}
where all the matrices are bosonic. The definitions for fermionic matrices can be similarly 
obtained by replacing the bosonic matrices with the fermionic ones.

The relations below follow  from the definition:
\begin{equation}
   Str(e^{ik\cdot A}B\cdot C) =Str(e^{ik\cdot A} C\cdot B),
\end{equation}
\begin{equation}
    Str(e^{ik\cdot A}B\cdot C\cdot D)=Str(e^{ik\cdot A}C\cdot B\cdot D)=Str(e^{ik\cdot A}C\cdot D\cdot B)=\cdot \cdot \cdot.
\end{equation}
That is, we can permute  matrices  in the  symmetrized trace. In particular, there are no ordering ambiguities in symmetrized trace. For fermionic matrices, 
an appropriate change of sign must be included.

Another useful equation related to the symmetrized trace is 
\begin{eqnarray}
  Str(e^{ik\cdot A}[ik\cdot A,B]C_1\cdot C_2 \cdot \cdot \cdot C_n) &=&
     -Str(e^{ik\cdot A}[C_1,B]\cdot C_2 \cdot \cdot \cdot C_n) \nonumber \\ 
    && -Str(e^{ik\cdot A}C_1\cdot [C_2,B]\cdot \cdot \cdot C_n)\nonumber \\
     && -\cdot \cdot \cdot \nonumber \\
      &&-Str(e^{ik\cdot A}C_1\cdot C_2 \cdot \cdot \cdot [C_n,B]) .
\end{eqnarray}

These relations above  are frequently used in this paper  to derive the vertex operators.

\subsubsection{The $\Xi$ term}
      Let us consider $Str(e^{ik\cdot A}\psi_{\alpha}\cdot\psi_{\beta})(\Gamma_0\Gamma_{\mu_1\cdot \cdot \cdot \mu_n})_{\alpha\beta}$ 
and $Str(e^{ik\cdot A}\{ \psi_{\alpha},\psi_{\beta} \})(\Gamma_0\Gamma_{\mu_1\cdot \cdot \cdot \mu_n})_{\alpha\beta}$,
 where $\psi$ is  an $n \times n$ Majorana-Weyl fermionic matrix. 
 Since hey  vanish if $n$ is even (because $\psi$ is  Weyl), we only need  to consider the case for 
odd $n$. 
     Note that $Str(e^{ik\cdot A}\psi_{\alpha}\cdot\psi_{\beta})$ is anti-symmetric in $\alpha$
 and $\beta$, whereas 
    $Str(e^{ik\cdot A}\{ \psi_{\alpha},\psi_{\beta} \})$ is symmetric. 
Due to
(\ref{symmetry_anti-symmetry_of_gamma_matrix}),
   $Str(e^{ik\cdot A}\psi_{\alpha}\cdot\psi_{\beta})(\Gamma_0\Gamma_{\mu_1\cdot \cdot \cdot \mu_n})_{\alpha\beta}$ is zero unless $n=3,7$, 
whereas  $Str(e^{ik\cdot A} \{ \psi_{\alpha},\psi_{\beta} \})$  $(\Gamma_0\Gamma_{\mu_1\cdot \cdot \cdot \mu_n})_{\alpha\beta}$ vanishes unless $n=1,5,9$.
$\Gamma_{\mu_1\cdot \cdot \cdot \mu_9}$ can be described by $\Gamma_{\nu_1}$ using the duality relation of gamma matrices(\ref{duality_of_gamma_matrix}),
and hence  reduces to $n=1$. 
In order to deal with  $\{\psi_{\alpha},\psi_{\beta}\}(\Gamma_0\Gamma_{\mu})_{\alpha\beta}$, we can make use of the equation of motion.  
For $n=5$, we introduce a new notation $\Xi$ as
\begin{equation}
  \Xi_{\mu_1\mu_2\mu_3\mu_4\mu_5} \equiv \{ \psi_{\alpha}, \psi_{\beta} \} (\Gamma_0 \Gamma_{\mu_1\mu_2\mu_3\mu_4\mu_5})_{\alpha\beta}.
\end{equation} 
%
Although unfamiliar in the literature,  it necessarily appears in the expressions of the vertex 
operators.

Finally, we give  a  useful Fierz identity related to $\{ \psi_{\alpha}, \psi_{\beta} \}$:
 \begin{eqnarray}
Str(e^{ik\cdot A} \bar{\psi}\cdot \Gamma_{\mu}\{\psi ,\bar{\psi}\}\Gamma_{\nu}\lambda)
&=& Str(e^{ik\cdot A} \psi_{\alpha}\cdot (\Gamma_0\Gamma_{\mu})_{\alpha\beta}\{\psi_{\beta},\psi_{\gamma}\} 
(\Gamma_0)_{\gamma\delta}(\Gamma_{\nu})_{\delta\epsilon}\lambda_{\epsilon})
\nonumber \\
&=& Str(e^{ik\cdot A}\frac{1}{16}\{ \psi_{\gamma},\psi_{\beta}\} (\Gamma_0 \Gamma^{\tau})_{\gamma\beta}\cdot
(\bar{\psi}\Gamma_{\mu}\Gamma_{\tau}\Gamma_{\nu}\lambda))
\nonumber \\
&&+Str(e^{ik\cdot A}\frac{1}{32}\cdot \frac{1}{5!}\Xi^{\tau_1\tau_2\tau_3\tau_4\tau_5}
\cdot(\bar{\psi}\Gamma_{\mu}\Gamma_{\tau_1\tau_2\tau_3\tau_4\tau_5}\Gamma_{\nu}\lambda)).\nonumber \\
\end{eqnarray}
If $k=0$ and $\mu =\nu$, the above formula becomes
\begin{equation}
tr \bar{\psi}\Gamma_{\mu}\{ \psi,\bar{\psi} \} \Gamma^{\mu}\lambda =
-tr \frac{1}{2}(\bar{\psi}\Gamma^{\tau}\lambda)\{ \bar{\psi}\Gamma_{\tau},\psi \}.
\end{equation}
\section{10D Vertex Operators}
\begin{eqnarray}
V^{\Phi}(A,\psi) &=& tr e^{ik\cdot A}\nonumber \\
V^{\tilde{\Phi}}(A,\psi) &=& Str e^{ik\cdot A}\bar{\psi}\nonumber \\
V^{B}_{\mu \nu}(A,\psi) &=& Str e^{ik\cdot A}\left(\frac{1}{16}k^{\rho}(\bar{\psi}\cdot \Gamma_{\mu\nu\rho}\psi)
               -\frac{i}{2}[A_{\mu},A_{\nu}]\right)\nonumber \\
V^{\Psi}_{\mu}(A,\psi)&=& Str e^{ik \cdot A}\left(-\frac{i}{12}k^{\rho}(\bar{\psi}\cdot \Gamma_{\mu\nu\rho}\psi)
                                 -2[A_{\mu},A_{\nu}]\right)\cdot\bar{\psi}\Gamma^{\nu}\nonumber \\
V^{h}_{\mu\nu}(A,\psi)&=&Str e^{ik\cdot A} \bigg(-\frac{1}{96}k^{\rho}k^{\sigma}\left(\bar{\psi}\cdot 
      {\Gamma_{\mu\rho}}^{\beta}\psi\right)\cdot\left(\bar{\psi}\cdot \Gamma_{\nu \sigma\beta} \psi \right) \nonumber \\
      &&-\frac{i}{4}k^{\rho}\bar{\psi}\cdot \Gamma_{\rho\beta(\mu}\psi \cdot {F_{\nu )}}^{\beta}+\frac{1}{2}\bar{\psi}
     \cdot \Gamma_{(\mu}[A_{\nu )},\psi]+2{F_{\mu}}^{\rho}\cdot F_{\nu\rho} \bigg) \nonumber \\
V^{A}_{\mu\nu\rho\sigma}(A,\psi)&=& Str e^{ik\cdot A} \bigg( \frac{i}{8\cdot 4!}k_{\alpha}k_{\gamma}
      (\bar{\psi}\cdot {\Gamma_{[\mu\nu}}^{\alpha}\psi)\cdot (\bar{\psi}\cdot {\Gamma_{\rho\sigma]}}^{\gamma}\psi) 
                                 +\frac{i}{3}\bar{\psi} \cdot \Gamma_{[\nu\rho\sigma}[\psi,A_{\mu]}]\nonumber \\
            &&+\frac{1}{4}F_{[\mu\nu}\cdot (\bar{\psi}\cdot {\Gamma_{\rho\sigma]}}^{\gamma}\psi)k_{\gamma}
          -iF_{[\mu\nu}\cdot F_{\rho\sigma]}\bigg)\nonumber \\
V^{\Psi^{c}}_{\mu}(A,\psi)&=&Str e^{ik\cdot A}\bigg( -\frac{i}{2\cdot 5 !}k^{\lambda}k^{\tau}(\bar{\psi}
             \cdot{\Gamma_{\mu \lambda}}^{\sigma}\psi)\cdot(\bar{\psi}\cdot \Gamma_{\nu\tau\sigma}\psi)\cdot 
           \bar{\psi}\Gamma^{\nu}\nonumber \\
            &&+\frac{1}{24}k^{\lambda}(\bar{\psi}\cdot \Gamma_{\lambda\mu\nu}\psi)\cdot \bar{\psi}
           \Gamma^{\nu}\Gamma_{\rho\sigma}\cdot F^{\rho\sigma}
            -\frac{1}{6}k^{\lambda}(\bar{\psi}\cdot \Gamma_{\lambda\alpha\beta}\psi)\cdot 
                            \bar{\psi}\Gamma^{\beta}\cdot {F^{\alpha}}_{\mu}\nonumber \\
              &&+\frac{i}{3}(\bar{\psi}\cdot \Gamma_{\mu}[A_{\nu},\psi])\cdot \bar{\psi}\Gamma^{\nu}
                +\frac{i}{3}(\bar{\psi}\cdot \Gamma_{\nu}[A_{\mu},\psi])\cdot \bar{\psi}\Gamma^{\nu}
         +\frac{i}{6}(\bar{\psi}\cdot \Gamma_{\alpha\beta\mu}\psi)\cdot [A^{\alpha},\bar{\psi}]\Gamma^{\beta}\nonumber \\
             &&-iF_{\mu\nu}\cdot F_{\rho\sigma}\cdot \bar{\psi}\Gamma^{\nu}\Gamma^{\rho\sigma}\bigg)\nonumber
             \end{eqnarray}
             \begin{eqnarray}
V^{B^{c}}_{\mu\nu}(A,\psi)&=& Str e^{ik\cdot A}\bigg( -\frac{1}{8 \cdot 6!}k^{\lambda}k^{\tau}k^{\alpha}(\bar{\psi}
        \cdot {\Gamma_{\mu\lambda}}^{\sigma}\psi)\cdot (\bar{\psi}\cdot \Gamma_{\gamma\tau\sigma} \psi)
                   \cdot (\bar{\psi}\cdot {\Gamma^{\gamma}}_{\alpha\nu} \psi) \nonumber \\
             && +\frac{i}{64}(\bar{\psi}\cdot\ks \Gamma_{\mu\alpha}\psi)\cdot F^{\alpha\beta}
             (\bar{\psi}\cdot \ks\Gamma_{\beta\nu}\psi)+\frac{i}{16\cdot 4!}(\bar{\psi}\cdot\ks 
                   \Gamma_{[\mu\alpha}\psi)\cdot(\bar{\psi}\cdot\ks \Gamma^{\alpha\sigma}\psi)\cdot F_{\sigma\nu]}\nonumber \\
             &&-\frac{1}{32}\bar{\psi}\cdot\Gamma_{[\mu}[A^{\sigma},\psi]\cdot (\bar{\psi}\cdot\ks\Gamma_{\sigma\nu]}\psi)
          -\frac{1}{64}(\bar{\psi}\cdot\ks\Gamma_{[\mu\alpha}\psi)\cdot \bar{\psi}\Gamma^{\alpha}[A_{\nu]},\psi] \nonumber \\
           &&+\frac{i}{4!\cdot 32}\Xi_{\mu\nu\alpha\beta\gamma}\cdot (\bar{\psi}\cdot \Gamma^{\alpha\beta\gamma}\psi)
            -\frac{i}{64}[A_{\alpha},F^{\alpha\tau}]\cdot (\bar{\psi}\cdot \Gamma_{\tau\mu\nu}\psi) \nonumber \\
           && +\frac{1}{64}(\bar{\psi}\cdot \Gamma_{\mu\nu\rho\sigma\lambda\tau}\ks \psi)\cdot F^{\rho\sigma}\cdot F^{\lambda\tau}
            +\frac{1}{16}(\bar{\psi} \cdot \Gamma_{\rho\sigma}\ks\psi)\cdot F^{\rho\sigma}\cdot F^{\mu\nu}\nonumber \\
           && -\frac{1}{8}(\bar{\psi}\cdot \Gamma_{\rho\sigma}\ks \psi ) \cdot F^{\mu\rho}\cdot F^{\nu\sigma}
            +\frac{1}{8}(\bar{\psi}\cdot \Gamma_{[\mu\sigma}\ks \psi ) \cdot F^{\sigma\alpha}\cdot F_{\alpha\nu]}\nonumber \\
          && -\frac{1}{32}(\bar{\psi}\cdot \Gamma^{\mu\nu}\ks \psi )\cdot  F^{\rho\sigma}\cdot F_{\sigma\rho} 
          +\frac{i}{4}\bar{\psi}\cdot \Gamma_{\mu\nu\alpha}[A_{\beta},\psi]\cdot F^{\alpha\beta}\nonumber \\
           && +\frac{i}{8}\bar{\psi}\cdot \Gamma_{\rho\sigma[\mu}[A_{\nu]},\psi]\cdot F^{\rho\sigma} 
           +\frac{i}{4}\bar{\psi}\cdot \Gamma_{(\mu}[A_{\rho)},\psi] \cdot{F^{\rho}}_{\nu} 
           -\frac{i}{4}\bar{\psi}\cdot \Gamma_{(\nu}[A_{\rho)},\psi]\cdot F^{\rho\mu} \nonumber \\
         && -iF_{\mu\rho}\cdot F^{\rho\sigma}\cdot F_{\sigma\nu}+\frac{i}{4}F_{\mu\nu}\cdot 
                            F^{\rho\sigma}\cdot F_{\sigma\rho}\bigg)\nonumber\\
V^{\tilde{\Phi}^c}(A,\psi) &=& Str e^{ik\cdot A} \bigg(\frac{1}{8!}(\bar{\psi}\cdot\Gamma^{\alpha\gamma}\ks \psi)
                            \cdot (\bar{\psi}\cdot \Gamma_{\gamma\delta}\ks\psi)\cdot(\bar{\psi}\cdot 
                                    \Gamma^{\delta\beta}\ks \psi)\cdot\bar{\psi} \Gamma_{\alpha\beta}\nonumber \\
                       &&\hskip -4em
                        -\frac{i}{2\cdot 5!}F^{\mu\alpha}\cdot (\bar{\psi}\cdot \ks\Gamma_{\alpha\beta}\psi)\cdot 
                                          (\bar{\psi}\cdot \ks \Gamma^{\beta\nu}\psi)\cdot \bar{\psi}\Gamma_{\mu\nu} 
                         +\cdot \cdot \cdot \nonumber \\
                    &&\hskip -4em  
                    \cdot \cdot \cdot -\frac{1}{8\cdot 4!}F^{\mu\nu}\cdot F^{\rho\sigma}(\bar{\psi}\cdot
                                    \Gamma_{\mu\nu\rho\sigma\lambda\alpha\beta}\psi)k^{\lambda}\cdot \bar{\psi}
                      \Gamma^{\alpha\beta} -\frac{1}{12}F^{\mu\alpha}\cdot F_{\alpha\beta}\cdot (\bar{\psi}\cdot \ks
                                           \Gamma^{\beta\nu}\psi)\cdot \bar{\psi}\Gamma_{\mu\nu} \nonumber \\
                      &&\hskip -4em
                      -\frac{1}{24}F^{\mu\alpha}\cdot (\bar{\psi}\cdot \ks\Gamma_{\alpha\beta}\psi)\cdot F^{\beta\nu}\cdot 
                         \bar{\psi}\Gamma_{\mu\nu} -\frac{1}{48}F^{\rho\sigma}\cdot (\bar{\psi} \cdot \ks
                      \Gamma_{\rho\sigma}\psi)\cdot F^{\mu\nu}\cdot \bar{\psi}\Gamma_{\mu\nu}+\cdot \cdot \cdot \nonumber \\
                    &&\hskip -4em
                     +\cdot \cdot \cdot +\frac{i}{24}\bar{\psi}\cdot \Gamma_{\mu\nu\rho\sigma\lambda\tau}F^{\mu\nu}\cdot
                            F^{\rho\sigma}\cdot F^{\lambda\tau}+i\bar{\psi}\cdot \Gamma^{\mu\nu}
                    (F_{\mu\rho}\cdot F^{\rho\sigma}\cdot F_{\sigma\nu}-\frac{1}{4}F^{\rho\sigma}\cdot F_{\sigma\rho}\cdot F_{\mu\nu})
                   \bigg)\nonumber \\
V^{\Phi^c}(A,\psi)&=&Str e^{ik\cdot A}\bigg(\frac{1}{8\cdot 8!}(\bar{\psi}\cdot \Gamma^{\alpha\gamma}\ks\psi)\cdot 
                              (\bar{\psi} \cdot \Gamma_{\gamma\delta}\ks \psi)
                      \cdot (\bar{\psi}\cdot \Gamma_{\delta\beta}\ks\psi)\cdot (\bar{\psi}\cdot \Gamma_{\alpha\beta}\ks \psi)
                            +\cdot \cdot\cdot 
                          \nonumber \\
                    &&\hskip -4em
                      \cdot \cdot \cdot+\frac{i}{48}(\bar{\psi}\cdot \Gamma_{\mu\nu\rho\sigma\lambda\tau}\ks\psi ) \cdot F^{\mu\nu}
                                 \cdot F^{\rho\sigma}\cdot F^{\lambda\tau} 
                 +[A_{\mu},\bar{\psi}]\cdot \Gamma_{\rho\sigma}\Gamma_{\nu}\psi \cdot F^{\mu\nu}\cdot F^{\rho\sigma}\nonumber \\
                   &&\hskip -4em
                     +\frac{i}{2}(\bar{\psi}\cdot \Gamma^{\mu\nu}\ks \psi)\cdot (F_{\mu\rho}\cdot F^{\rho\sigma}
                            \cdot F_{\sigma\nu}-\frac{1}{4}F^{\rho\sigma}\cdot F_{\sigma\rho}\cdot F_{\mu\nu})  \nonumber \\
                       &&\hskip -4em
                         -(F_{\mu\nu}\cdot F^{\nu\rho}\cdot F_{\rho\sigma}\cdot F^{\sigma\mu}-\frac{1}{4}F_{\mu\nu}\cdot F^{\nu\mu}
                     \cdot F_{\rho\sigma} \cdot F^{\sigma\rho})\bigg).\nonumber
\end{eqnarray}

\end{document}